\documentclass[preprint,showpacs,preprintnumbers,amsmath,amssymb,superscriptaddress,nofootinbib]{revtex4}
\usepackage{graphicx,color}
\usepackage{amsmath,amssymb}
\usepackage{url}
\usepackage{epstopdf}

\newcommand{\be}{\begin{equation}}
	\newcommand{\ee}{\end{equation}}
\newcommand{\bea}{\begin{eqnarray}}
	\newcommand{\eea}{\end{eqnarray}}
\newcommand{\nn}{\nonumber}
\newcommand{\crn}{\nonumber \\}

\newcommand{\fr}{\frac}

\newcommand{\bc}{\begin{center}}
	\newcommand{\ec}{\end{center}}

\newcommand {\ba}{\begin{array}}
	\newcommand {\ea}{\end{array}}
\newcommand{\ben}{\begin{enumerate}}
	\newcommand{\een}{\end{enumerate}}

\usepackage{graphicx,color}

\usepackage{bm}
\usepackage{dcolumn}

\begin{document}
	
	\title{Decays $h\to e_ae_b$, $e_b\to e_a\gamma$, and $(g-2)_{e,\mu}$ in a 3-3-1 model with  inverse seesaw neutrinos}

	\author{T.T. Hong}\email{tthong@agu.edu.vn}
	\affiliation{An Giang University, VNU - HCM, Ung Van Khiem Street,
		Long Xuyen, An Giang 88000, Vietnam}
	\author{N.H.T. Nha}\email{nhtnha95@gmail.com }
	\affiliation{Department of Physics, Can Tho University,
		3/2 Street, Can Tho, Vietnam}

	\author{T. Phong Nguyen}\email{thanhphong@ctu.edu.vn}
	\affiliation{Department of Physics, Can Tho University,
		3/2 Street, Can Tho, Vietnam}
	\author{L.~T.~T.~Phuong}
	\email{lttphuong@agu.edu.vn}
	\affiliation{An Giang University, VNU - HCM, Ung Van Khiem Street,
		Long Xuyen, An Giang 88000, Vietnam} 
	\author{L.T. Hue\footnote{Corresponding author} }\email{lethohue@vlu.edu.vn}
	\affiliation{Subatomic Physics Research Group, Science and Technology Advanced Institute, Van Lang University, Ho Chi Minh City 70000, Vietnam}
	\affiliation{Faculty of Applied Technology, School of Engineering and Technology, Van Lang University, Ho Chi Minh City 70000, Vietnam}

	\begin{abstract}
		We  will show that the 3-3-1 model with new heavy right handed neutrinos  as $SU(3)_L$ singlets can explain simultaneously the lepton flavor violating decays of the SM-like Higgs boson, charged lepton flavor violating decays $e_b\rightarrow e_a\gamma$, and the electron $(g-2)_e$ anomalies under recent experimental data. The discrepancy  of $(g-2)_{\mu}$ predicted by the model under consideration and that of the standard model  can reach $10^{-9}$.  The decay rates of the standard model-like Higgs boson $h\to \tau e, \tau\mu$ can reach the values of $\mathcal{O}(10^{-4})$. 	\end{abstract} 
	\maketitle
\section{\label{intro} Introduction}
\allowdisplaybreaks

The experimental evidence of neutrino oscillation \cite{Super-Kamiokande:1998kpq, Super-Kamiokande:2001ljr, Super-Kamiokande:2001bfk, SNO:2002tuh, SNO:2002hgz}  confirms  that the lepton flavor number is violated in the neutral lepton sector. This is a great motivation to search for many lepton flavor violating (LFV) processes, namely the promoting ones we will focus on in this work are the LFV decays of the charged leptons $e_b\to e_a\gamma$ and the standard model-like (SM-like) Higgs boson  (LFVH) $h\to e^\pm_ae^\mp_b$. 
 The charged lepton flavor violating  (cLFV) decays  $e_b\rightarrow e_a\gamma$ are  constrained by  experiments as follows~\cite{BaBar:2009hkt, MEG:2016leq}:
\begin{align}
	\label{eq_ebagaex}
	\mathrm{Br}(\tau\rightarrow \mu\gamma)&<4.4\times 10^{-8}, \; 
	\mathrm{Br}(\tau\rightarrow e\gamma) <3.3\times 10^{-8}, \;
	\mathrm{Br}(\mu\rightarrow e\gamma) < 4.2\times 10^{-13}.
\end{align}
 Upcoming sensitivities  will be  orders  of  $10^{-9}$ and $10^{-14}$ for decays $\tau\to \mu\gamma,e\gamma$ \cite{Belle-II:2018jsg,  Aushev:2010bq}   and $\mu\to e\gamma$ \cite{MEGII:2018kmf}, respectively. LFVH decays  have been investigated in many models beyond the standard model (BSM). On the other hand, the latest experimental constraints  are: Br$(h\to \tau^\pm \mu^{\mp})<2.5\times 10^{-3}$ \cite{CMS:2017con}, Br$(h\to \tau^\pm e^{\mp})<4.7\times 10^{-3}$ \cite{ATLAS:2019pmk}, and Br$(h\to\mu^\pm e^{\mp})<6.1\times 10^{-5}$ \cite{ATLAS:2019xlq}.  The future  experimental sensitivities may be  $1.4\times 10^{-4}$,  $1.6\times 10^{-4}$, and $1.2 \times 10^{-5}$, respectively \cite{Qin:2017aju}. The small upper bounds of the cLFV branching rates prefer the explanation that they come from loop corrections relevant to LFV sources, including ones available in the neutral lepton sector.  For models consisting of  these necessary tree level couplings to accommodate neutrino oscillation data such as the Zee model~\cite{Zee:1980ai}, constraints on the LFV sources such as Yukawa couplings are very strict \cite{Herrero-Garcia:2017xdu, Barman:2021xeq}. Therefore, new scalar masses must not be heavier than 300 GeV in order to explain successfully the recent $(g-2)$ data~\cite{Barman:2021xeq, Sabatta:2019nfg}, while the LFVH decay rates are small \cite{Herrero-Garcia:2017xdu, Vicente:2019ykr}. 

To explain the neutrino oscillation data, the BSMs with the general seesaw(GSS)  mechanism also result in LFV decays. But the  versions adding only heavy seesaw neutrinos type-I predict  suppressed LFV rates that are much smaller than the  upcoming experimental sensitivities \cite{Arganda:2004bz, Marcano:2019rmk}. In contrast,  the models with only new inverse seesaw (ISS) neutrinos can predict  large LFV  rates. In addition, LFVH rates may be large   in the regions satisfying  constraints of Br$(e_b\to e_a\gamma)$ \cite{Ilakovac:1999md, Arganda:2014dta, Arganda:2016zvc, Thao:2017qtn}.  On the other  hand,  LFVH rates may be smaller when other  constraints   are considered  \cite{Hernandez-Tome:2020lmh, Nguyen:2020ehj}. In the supersymmetric (SUSY) versions of these models with new LFV sources from superparticles, LFVH rates may reach large order of $\mathcal{O}(10^{-5})$ \cite{Brignole:2003iv, Brignole:2004ah, Diaz-Cruz:2002ezb, Arganda:2004bz, Giang:2012vs,  Arana-Catania:2013xma, Hue:2015fbb,Arganda:2015uca, Arganda:2015naa, Zeleny-Mora:2021tym}.   LFVH decays were also addressed  with other experimental data in many other non-SUSY extensions of the SM \cite{Aoki:2016wyl, Cheung:2015yga, Chen:2016lsr, Yang:2016hrh, Guo:2016ixx,  Huitu:2016pwk,  Chang:2016ave, Altmannshofer:2015esa, Omura:2015xcg, Lami:2016mjf, Das:2015zwa, Crivellin:2015mga, Campos:2014zaa, Omura:2015nja, deLima:2015pqa, Heeck:2014qea, Dorsner:2015mja, He:2015rqa, Dery:2014kxa, Celis:2013xja, Falkowski:2013jya, Harnik:2012pb,BhupalDev:2012zg, Goudelis:2011un, Diaz-Cruz:1999sns,  Korner:1992zk, Pilaftsis:1991ug, Pilaftsis:1992st, Blankenburg:2012ex,CarcamoHernandez:2014wdl}.   Many BSM predict that the strong constraints of cLFV decay rates Br$(e_b\to e_a\gamma)$  give small  LFVH ones, or suppressed  $(g-2)_{\mu}$. 

 Unless there is some specific condition of the appearance of very light new bosons, the above cLFV constraints will result in small new one-loop contributions  to anomalous magnetic moments (AMMs) of charged leptons $(g-2)_{e_a}/2\equiv a_{e_a}$, in contrast with recent  experimental data. Namely,  the $4.2\sigma$  deviation between  standard model (SM) prediction \cite{Aoyama:2020ynm}, combined contributions from previous works \cite{Aoyama:2012wk, Aoyama:2019ryr,  Czarnecki:2002nt,  Gnendiger:2013pva, Davier:2017zfy, Keshavarzi:2018mgv, Colangelo:2018mtw, Hoferichter:2019mqg, Davier:2019can, Keshavarzi:2019abf, Kurz:2014wya, Melnikov:2003xd, Masjuan:2017tvw, Colangelo:2017fiz, Hoferichter:2018kwz, Gerardin:2019vio, Bijnens:2019ghy, Colangelo:2019uex, Blum:2019ugy, Colangelo:2014qya, Pauk:2014rta, Danilkin:2016hnh, Jegerlehner:2017gek, Knecht:2018sci, Eichmann:2019bqf, Roig:2019reh},  and muon experiments  \cite{Muong-2:2006rrc, Muong-2:2021ojo} is   
\begin{equation} 
	\label{eq_damu}
	\Delta a^{\mathrm{NP}}_{\mu}\equiv  a^{\mathrm{exp}}_{\mu} -a^{\mathrm{SM}}_{\mu} =\left(2.51 \pm 0.59\right)  \times 10^{-9}.
\end{equation}
This result is slightly inconsistent with the latest one, which calculated  the hadronic vacuum polarization for the SM prediction based on the lattice QCD approach,  giving a combined value  reported in Refs.  \cite{Borsanyi:2020mff, Davier:2019can,Keshavarzi:2019abf}  closer to the experimental data. This value was shown to fit  with  other experimental data such as global  electroweak fits \cite{Crivellin:2020zul, Keshavarzi:2020bfy, Colangelo:2020lcg}.   

Regarding  the electron anomaly, a $1.6\sigma$ discrepancy between SM and experiment was reported    \cite{Morel:2020dww}
\begin{equation}\label{eq_dae}
	\Delta a^{\mathrm{NP}}_{e}\equiv  a^{\mathrm{exp}}_{e} -a^{\mathrm{SM}}_{e} = \left( 4.8\pm 3.0\right) \times 10^{-13}.
\end{equation}  
The recent studies of cLFV decays in the regions satisfying the AMM data were done in some specific models such as SUSY  with largest Br$(h\to \tau\mu)\sim \mathcal{O}(10^{-4})$ \cite{Zhang:2021nzv}. Other  BSM containing  leptoquarks  can explain the  large $\Delta a^{\mathrm{NP}}_{\mu}\sim  \mathcal{O}(10^{-9}) $ \cite{Baek:2015mea}.

Recent work has discussed an extension of the  3-3-1 with right-handed neutrinos \cite{Foot:1994ym, Long:1996rfd, Long:1995ctv}, named the 3-3-1 model with inverse seesaw neutrinos (331ISS) \cite{Hue:2021xap},  with the aim of giving an explanation of both the $(g-2)_{\mu}$ data and the neutrino oscillation data through the ISS  mechanism.  The model  needs new $SU(3)_L$ gauge singlets including  three neutral leptons $X_{aR}$ and a new singly charged Higgs  boson $h^\pm$ to  accommodate all the experimental data of neutrino oscillation,   the cLFV bounds  in Eq. \eqref{eq_ebagaex} and  the $\Delta a_{\mu}$ in $1\sigma$ deviation given in Eq. \eqref{eq_damu}.  Although  cLFV  and/or LFVH decays   were investigated previously with promoting predictions for the 331ISS  \cite{Zhang:2015csm, Nguyen:2018rlb, CarcamoHernandez:2020pnh, Hung:2021fzb}, the AMM data was  not included. Our aim in this work is filling this gap.  We  note that other 3-3-1 models \cite{Montero:1992jk, Singer:1980sw,Frampton:1992wt, Pisano:1992bxx} constructed previously can accommodate the $(g-2)_{\mu}$ data only when they are extended such as adding new vector-like fermions, or/and scalars \cite{Lindner:2016bgg, DeJesus:2020yqx, deJesus:2020ngn, Hernandez:2021xet, Hue:2021zyw}. But none of them paid attention to the correlations between LFVH decays and  $(g-2)_{e_a}$ anomalies.

Our paper is organized as follows. In Sec.~\ref{model} we  discuss the necessary ingredients of a 331ISS model for studying LFVH decays  and how the ISS mechanism works to generate active neutrino masses and mixing  consistent with current experimental data. In Sec.~\ref{sec_coupling} we present all couplings needed to determine the one-loop contributions to  the LFVH decay amplitudes  of the SM-like Higgs boson, cLFV decays, and  $(g-2)_{e_a}$. In section \ref{sec_numerical}, we provide detailed numerical illustrations and discussions. Section \ref{conclusion} contains our conclusions. Finally, the appendix  lists all of the analytic formulas expressing one-loop contributions to LFVH decay amplitudes  calculated in the unitary gauge.

\section{\label{model} The 331ISS model for tree-level neutrino masses}
\subsection{Particle content and lepton masses} 
We summarize the particle content of the 331ISS model in this section. We ignore  the quark sector irrelevant in our  work, which was  discussed previously \cite{Boucenna:2015zwa, Chang:2006aa}. We also ignore many detailed calculations presented in Ref. \cite{Hue:2021xap}. The electric charge operator defined by  the gauge group  $SU(3)_L\times U(1)_X$ is $Q=T_3-\frac{1}{\sqrt{3}}T_8+X$, where $T_{3,8}$ are diagonal $SU(3)_L$ generators. Each  lepton family consists of  an  $SU(3)_L$ triplet $L_{aL}= (\nu_a,~e_a, N_a)_L^T\sim (3,-\frac{1}{3})$ and a right-handed charged lepton  $e_{aR}\sim (1,-1)$ with $a=1,2,3$. The 331ISS model contains  three  neutral leptons  $X_{aR}\sim (1,0)$, $a=1,2,3$, and a singly charged Higgs boson $\sigma^\pm\sim (1,\pm1)$. There are  three Higgs triplets  $\rho=(\rho^+_1,~\rho^0,~\rho^+_2)^T\sim (3,\frac{2}{3})$,
$\eta=(\eta_1^0,~\eta^-,\eta^0_2)^T\sim (3,-\frac{1}{3})$, and $\chi=(\chi_1^0,~\chi^-,\chi^0_2)^T\sim (3,-\frac{1}{3})$.  The vacuum expectation values (vev) for generating all tree-level quark masses and leptons are $\langle\rho \rangle=(0,\,\frac{v_1}{\sqrt{2}},\,0)^T$, $\langle \eta \rangle=(\frac{v_2}{\sqrt{2}},\,0,\,0)^T$ and $\langle \chi \rangle=(0,\,0,\,\frac{w}{\sqrt{2}})^T$. Two neutral Higgs components have  zero vevs because of  their  non-zero generalized lepton numbers~\cite{Hue:2021xap} corresponding to a new global symmetry $U(1)_\mathcal{L}$ \cite{Chang:2006aa}.

In the 331ISS, nine gauge bosons  get masses through the covariant kinetic Lagrangian of the Higgs  triplets, $\mathcal{L}^{H}=\sum_{H=\chi,\eta,\rho} \left(D_{\mu}H\right)^{\dagger}\left(D^{\mu}H\right)$,  where 
$D_\mu  = \partial _\mu  - i g  \sum_{a=1}^{8}{W}_\mu ^a{T^a} - i{g_X}{T^9}X {X_\mu }$,  $a=1,2,..,8$, and   $T^9 \equiv \frac{I_3}{\sqrt{6}}$ and $\frac{1}{\sqrt{6}}$ for (anti)triplets and singlets \cite{Buras:2012dp}.  There are two pairs of singly charged gauge bosons, denoted as    $W^{\pm}$  and $Y^{\pm}$, defined as
\begin{align}
	W^{\pm}_{\mu}&=\frac{W^1_{\mu}\mp i W^2_{\mu}}{\sqrt{2}},\; Y^{\pm}_{\mu}=\frac{W^6_{\mu}\pm i W^7_{\mu}}{\sqrt{2}},\; \;  
 \label{singlyG}
\end{align}
with the respective  masses  $m_W^2=\frac{g^2}{4}\left(v_1^2+v_2^2\right)$ and $	m_Y^2=\frac{g^2}{4}\left(w^2+v_1^2\right)$. The   breaking pattern of the model is $SU(3)_L\times U(1)_X\to SU(2)_L\times U(1)_Y\to U(1)_Q$, leading to the matching condition that  $W^{\pm}$ are  the SM gauge bosons. As a consequence,  we have     
\begin{align} \label{eq_SMi}
	v_1^2+v_2^2\equiv v^2=(246 \mathrm{GeV})^2,\quad  \frac{g_X}{g}= \frac{3\sqrt{2}s_W}{\sqrt{3-4s^2_W}},\quad  gs_W=e,
\end{align}
where $e$ and $s_W$ are, respectively, the electric charge and sine of the Weinberg angle. Similarly to the Two Higgs Doublet Models (2HDM),  we use the parameter 
\begin{equation}\label{eq_tbeta}
	t_{\beta}\equiv \tan\beta=\frac{v_2}{v_1},
\end{equation}
which leads to $v_1=vc_{\beta}$ and $v_2=vs_{\beta}$.

The Yukawa Lagrangian generating lepton masses is:
\begin{align}
	\label{eq_Lylepton}
	\mathcal{L}^{\mathrm{Y}}_l =&-h^e_{ab}\overline{L_{a}}\rho e_{bR}+
	h^{\nu}_{ab} \epsilon^{ijk} \overline{(L_{a})_i}(L_{b})^c_j\rho^*_k 	
	%
	- y^{\chi}_{ba}\overline{ X_{bR}}\chi^{\dagger} L_{a} -\frac{1}{2} (\mu_{X})_{ab}\overline{X_{aR}} \left(X_{bR}\right)^c
	\crn& -Y^{\sigma}_{ab} \overline{(X_{aR})^c}e_{bR}\sigma^{+} +\mathrm{H.c.},
\end{align}
where $a,b=1,2,3$. 
The first term generates charged lepton masses as $	m_{e_a}\equiv \frac{h^e_{ab}v_1}{\sqrt{2}} \delta_{ab}$,  with the assumption that  the flavor states are also physical. 

In the basis $n'_{L}=(\nu_{L}, N_{L}, (X_R)^c)^T$,   Lagrangian in Eq.  \eqref{eq_Lylepton} generates  a   neutrino mass term written in terms of   the total $9\times 9$ mass matrix  consisting of nine $3\times3$ sub-matrices  \cite{Nguyen:2018rlb}, namely 
\begin{align}
	-{\mathcal{L}}^{\nu}_{\mathrm{mass}}=\frac{1}{2}\overline{(n'_L)^c}\mathcal{M}^{\nu}n'_L  +\mathrm{H.c.}, \,\mathrm{ where }\quad \mathcal{M}^{\nu}=\begin{pmatrix}
		\mathcal{O}_{3}	& m^T_{D} & 	\mathcal{O}_{3}\\
		m_{D}	&\mathcal{O}_{3}	  & M^T_R \\
		\mathcal{O}_{3}& M_R& \mu_X
	\end{pmatrix},  \label{Lnu1}
\end{align}
where $(n'_{L})^c=((\nu_L)^c, (N_L)^c, X_R)^T$,  $(M_R)_{ab}\equiv y^{\chi}_{ab}\frac{w}{\sqrt{2}}$, and $(m^T_D)_{ab}= -(m_D)_{ab}\equiv \sqrt{2}h^{\nu}_{ab}v_1$  with $a,b=1,2,3$. The matrix  $\mu_{X}$  in Eq.~\eqref{eq_Lylepton}  is symmetric,  and can  be considered  as a diagonal matrix without loss of generality. 

The mass matrix  $\mathcal{M}^{\nu}$ is   diagonalized by a $9\times9$ unitary matrix $U^{\nu}$,
\begin{align}
	U^{\nu T}\mathcal{M}^{\nu}U^{\nu}=\hat{M}^{\nu}=\mathrm{diag}(m_{n_1},m_{n_2},..., m_{n_{9}})=\mathrm{diag}(\hat{m}_{\nu}, \hat{M}_N), \label{diaMnu}
\end{align}
where $m_{n_i}$ ($i=1,2,\dots 9$) are  masses corresponding to the  physical states $n_{iL}$. The two mass matrices    $\hat{m}_{\nu}=\mathrm{diag}(m_{n_1},\;m_{n_2},\;m_{n_3})$  and   $\hat{M}_N$ $=\mathrm{diag}(m_{n_4},\;m_{n_5},...,\;m_{n_{9}})$  consist of  the masses of the  active  $n_{aL}$ ($a=1,2,3$)  and extra neutrinos  $n_{IL}$ ($I=1,2,..,6$), respectively.  
The following approximation solution of $U^{\nu}$ is valid for any specific seesaw mechanisms,
\begin{align}
	\label{eq_Unu}
	U^{\nu}= \Omega \left(
	\begin{array}{cc}
		U_{\mathrm{PMNS}} & \mathcal{O}_{3\times6}\\
		\mathcal{O}_{6\times3} & V \\
	\end{array}
	\right), \;\; \Omega \simeq 
	\left(
	\begin{array}{cc}
		I_3-\frac{1}{2}RR^{\dagger} & R \\
		-R^\dagger &  I_6-\frac{1}{2}R^{\dagger} R\\
	\end{array}
	\right),
\end{align}
where $R$, $V$ are $3\times6$,  $3\times6$ matrices, respectively.  All entries of  $R$ must  satisfy $|R_{aI}|\ll$1, so that  all ISS relations can be derived perturbatively.  

The relations between the flavor  and mass eigenstates are
\begin{equation}
	n'_L=U^{\nu} n_L, \quad  \mathrm{and} \; (n'_L)^c=U^{\nu*}  (n_L)^c \equiv U^{\nu*}  n_R, \label{Nutrans}
\end{equation}
where $n_L\equiv(n_{1L},n_{2L},...,n_{9L})^T$, and the Majorana states are $n_i=(n_{iL},\;n_{iR})^T$. 

The ISS relations are 
\begin{align}
	R^*_2&=m_D^TM^{-1}_R,  \; 
	R^*_1 = -R^*_2 \mu_X\left( M^T_R\right)^{-1}\simeq \mathcal{O}_{3\time3},  \label{eq_R12}
	\\  m_{\nu} &=R_2^*\mu_XR_2^{\dagger} = U^*_{\mathrm{PMNS}} \hat{m}_{\nu}U^\dagger_{\mathrm{PMNS}} =m_D^TM^{-1}_R\mu_X\left( M^{-1}_R\right)^Tm_D,  \label{eq_mnu}
	\\ V^*\hat{M}_NV^{\dagger}&=M_N +\frac{1}{2} M_N R^{\dagger}R +	\frac{1}{2}R^{T}R^*M_N.  \label{eq_MN}
\end{align} 
From experimental data of $m_{\nu}$, we can determine all independent parameters in $m_D$  and three entries of $M^{-1}\equiv M^{-1}_R\mu_X\left( M^{-1}_R\right)^T$ \cite{Boucenna:2015zwa, Nguyen:2018rlb}. Namely,  the Dirac mass matrix has the antisymmetric form 
\begin{equation} \label{eq_mD}
	m_D= ze^{i\alpha_{23}}\times  \tilde{m}_D,\; 
\end{equation}
where $ \alpha_{23}\equiv \arg[h^{\nu}_{32}]$,   $\tilde{m}_D$ is an antisymmetric matrix with $(\tilde{m}_D)_{23}=1$, and 
 \begin{equation}\label{eq_zdef}
	z=\sqrt{2}v_1 |h^{\nu}_{32}|=\sqrt{2} v_1\,|h^{\nu}_{23}|\equiv z_0c_{\beta}	
\end{equation}
is a positive and real parameter.  Eq. \eqref{eq_mnu} gives $\left( m_{\nu}\right)_{ij}=\left[m_D^TM^{-1}m_D\right]_{ij}$ for all $i,j=1,2,3$, leading to six independent equations. Solving three of them with $i\neq j$,  the non-diagonal entries of $M^{-1}$ are functions of $M^{-1}_{ii}$ and $x_{12,13}$.   Inserting these functions into the three remaining relations with $i=j$, we obtain 
\begin{align}
	\label{eq_xijM}	
	(\tilde{m}_D)_{32}=\frac{(m_{\nu })_{13}^2-\left(m_{\nu })_{11}\right. \left(m_{\nu})_{33}\right.}{\left(m_{\nu })_{13}\right. \left(m_{\nu })_{23}\right.-\left(m_{\nu })_{12}\right. \left(m_{\nu
		})_{33}\right.},\; 
	 \; (\tilde{m}_D)_{21}= \frac{\left(m_{\nu })_{12}\right. \left(m_{\nu })_{13}\right.-\left(m_{\nu })_{11}\right. \left(m_{\nu })_{23}\right.}{\left(m_{\nu })_{13}\right. \left(m_{\nu })_{23}\right.-\left(m_{\nu })_{12}\right.
		\left(m_{\nu })_{33}\right.},
\end{align}
and $\mathrm{Det}[m_{\nu}]=0$. From $M^{-1}=M^{-1}_R\mu_X\left( M^{-1}_R\right)^T$  we  derive that three parameters  of the matrix $\mu_{X}$ as certain but lengthy functions of $(ze^{i\alpha_{23}})$, all entries of  $M_R$ and $m_{\nu}$. While $m_{\nu}$ are fixed by  experiments, all entries of $M_R$  are free parameters.  We will fix $\alpha_{23}=0$, because it is absorbed into the $\mu_X$.  

In the limit that $|R_2|\ll1$, the heavy neutrino masses can be determined  approximately based on Eq. \eqref{eq_MN}, namely 
\begin{align}
	V^*\hat{M}_NV^{\dagger}\simeq M_N.  \label{eq_MNa}
\end{align}
We  define the reduced matrix $	M_R\equiv z  \tilde{M}_R, \;  \left(\tilde{M}_R\right)_{ij}\equiv k_{ij}$, 
 provided that  $R_2^*=-\tilde{m}_D/\tilde{M}_R$.  The matrix  $M_R$ is always diagonalized by two unitary transformations $V_{L,R}$ \cite{Dreiner:2008tw}: 
\begin{align}
	V_L^TM_RV_R= z\times \hat{k}=z \times \mathrm{diag}(\hat{k}_1,\;\hat{k}_2,\;\hat{k}_3), \label{eq_MRd}
\end{align}
where all $\hat{k}_{1,2,3}$ are always positive and $\hat{k}_a\gg1$ so that all ISS relations are valid. Therefore,   $M_R$ is expressed in terms of   $\hat{k}$ and  $V_{L,R}$. Then the matrix $V$ in Eq. \eqref{eq_MN} can be found approximately as follows
\begin{align}
	V=\frac{1}{\sqrt{2}} \begin{pmatrix}
		V_R	&  iV_R\\
		V_L & -iV_L
	\end{pmatrix} \to V^TM_NV= z\times \begin{pmatrix}
		\hat{k}	& \mathcal{O}_{3\times3}\\
		\mathcal{O}_{3\times3} & \hat{k}
	\end{pmatrix}.  \label{eq_V0}
\end{align}
As a consequence,  for any qualitatively estimations we use the approximation    that heavy neutrinos masses are  $m_{n_{a+3}}=m_{n_{a+6}}\simeq z \hat{k}_a$ with a=1,2,3; $R_1\simeq \mathcal{O}_3$;  and
\begin{align} \label{eq_Unu1}
	U^{\nu}\simeq \begin{pmatrix} 
		\left( I_3 -\frac{1}{2} R_2R^\dagger_2\right) U_{\mathrm{PMNS}}&  \frac{1}{\sqrt{2}}R_2 V_L& \frac{-i}{\sqrt{2}}R_2 V_L \\
		\mathcal{O}_3& \frac{V_R}{\sqrt{2}} &  \frac{iV_R}{\sqrt{2}}\\
		-R^\dagger_2U_{\mathrm{PMNS}}& \left(I_3 -\frac{R^\dagger_2R_2}{2} \right)\frac{V_R}{\sqrt{2}}  
		& \left(I_3 -\frac{R^\dagger_2R_2}{2}\right)\frac{-iV_R}{\sqrt{2}}
	\end{pmatrix}. 
\end{align}
We have checked and confirmed that the above approximations   give numerical results consistent with those  discussed in Ref. \cite{Hue:2021xap}.  Therefore, these  approximate formulas will be used in this work.  $m_{\nu}$  is chosen as  the input with $3\sigma$ neutrino oscillation data to fix $\tilde{m}_D$.  The free parameters $z_0$ and $\hat{k}_{1,2,3}$, $V_{R}$ will be scanned  in the valid ranges to construct the total neutrino mixing matrix $U^{\nu}$ defined in Eq. \eqref{eq_Unu1}.  Because 
\begin{equation}\label{eq_R2VL}
	R_2V_L=\tilde{m}^{\dagger}_DV^*_R\hat{k}^{-1},	\quad R_2R_2^{\dagger}=\tilde{m}^{\dagger}_DV_R\hat{k}^{-2}\tilde{m}_D,
\end{equation}
which  do not depend explicitly  on  $V_L$,  it affects weakly on all relevant  processes.  We will  fix $V_L=I_3$ from now on.

Lagrangian for quark masses were discussed previously~\cite{Boucenna:2015zwa, Chang:2006aa}. Here, we just recall  that the Yukawa couplings of the top quark must satisfy  the perturbative limit $h^u_{33}<\sqrt{4\pi}$, leading to a lower bound of $v_2$:  $v_2 >\frac{\sqrt{2}m_t}{\sqrt{4\pi}}$. Combined with the relations in Eqs.~\eqref{eq_SMi} and ~\eqref{eq_tbeta}, the lower bound of $t_{\beta}$ is $t_{\beta}\ge0.3$. The upper bound of $t_{\beta}$ can be derived from the tau mass, $m_{\tau}=h^3_{33}\times vc_{\beta}\sqrt{2}\to h^3_{33}=m_{\tau}\sqrt{2}/(v c_{\beta})<\sqrt{4\pi}$, leading to  a rather weak upper bound $t_{\beta}=\sqrt{1/c^2_{\beta}-1}\leq 346$. 

\subsection{Higgs bosons}
The Higgs potential  used here respect the new lepton number defined in Ref.~\cite{Chang:2006aa}, namely 
\begin{align}
	\label{eq_Vh}
	V_{h}&= \sum_{S} \left[ \mu_S^2 S^{\dagger}S +\lambda_S \left(S^{\dagger}S\right)^2 \right]  + \lambda_{12}(\eta^{\dagger}\eta)(\rho^{\dagger}\rho)
	+\lambda_{13}(\eta^{\dagger}\eta)(\chi^{\dagger}\chi)
	+\lambda_{23}(\rho^{\dagger}\rho)(\chi^{\dagger}\chi)  \crn
	& +\tilde{\lambda}_{12} (\eta^{\dagger}\rho)(\rho^{\dagger}\eta) 
	+\tilde{\lambda}_{13} (\eta^{\dagger}\chi)(\chi^{\dagger}\eta)
	+\tilde{\lambda}_{23} (\rho^{\dagger}\chi)(\chi^{\dagger}\rho) +\sqrt{2} \omega f\left(\epsilon_{ijk}\eta^i\rho^j\chi^k +\mathrm{h.c.} \right)
	\crn & + \sigma^+\sigma^- \left[ \mu^2_{\sigma} +  \sum_{S} \lambda^{\sigma}_S S^{\dagger}S \right] + \left[ f_{\eta}(\rho^{\dagger}\eta) \sigma^++ f_{\chi}(\rho^{\dagger}\chi)\sigma^+  + \mathrm{h.c.}\right], 
\end{align}
where $f$ is a dimensionless parameter, $f_{\eta,\chi}$ are mass dimensional, $S=\eta,\rho,\chi$.  These three trilinear couplings  softly break  the general lepton number $\mathcal{L}$. For simplicity, we fix $f_\chi=0$ by applying a suitable discrete symmetry. The last line in Eq. \eqref{eq_Vh} contains all additional terms couplings with new charged Higgs singlets compared with Higgs potential considered in previous works \cite{Hue:2021xap}. They do not affect the squared mass matrices of both neutral CP-odd and CP-even Higgs bosons. The minimum conditions of the Higgs potential  as well as the identification of the SM-like Higgs boson have previously been discussed in detailed ~\cite{Ninh:2005su, Hue:2015fbb}, hence we just list the necessary results here.  The model contains three pairs of singly charged Higgs bosons $h^{\pm}_{1,2,3}$ and two Goldstone bosons $G^{\pm}_{W,Y}$ of the singly charged gauge bosons $W^{\pm}$ and $Y^{\pm}$, respectively. In the limit of $f_\eta=0$, the singly charged Higgs masses are  $m^2_{h^{\pm}_1}=  \left( \frac{\tilde{\lambda}_{12}v^2}{2} +\frac{fw^2}{s_{\beta}c_{\beta}} \right)$,  $m^2_{h^{\pm}_2}= (v^2 c_{\beta}^2+w^2)\left(\frac{\tilde{\lambda}_{23}}{2}  +ft_{\beta}\right)$ and  $m^2_{G^{\pm}_{W}}=m^2_{G^{\pm}_{Y}}=0$  \cite{Ninh:2005su}. The mass of the  Higgs singlet   $\sigma\equiv h^\pm_3$ is a function of $\mu^2_s$ and $\lambda^{\sigma}_S$.  With $f_{\eta}\neq0$ considered in this work, the relations between the original and mass eigenstates of the charged Higgs bosons  are
\begin{eqnarray}
	\left( \begin{array}{c}
		\eta^\pm\\
		\rho_1^\pm \\
		\sigma^\pm
	\end{array} \right)=\begin{pmatrix}
		-s_{\beta}& c_{\alpha} c_{\beta} &  s_{\alpha} c_{\beta}\\
		c_{\beta}& c_{\alpha} s_{\beta} &  s_{\alpha} s_{\beta}\\
		0 & -s_{\alpha}   & c_{\alpha} 
	\end{pmatrix} \left(\begin{array}{c}
		G_W^\pm
		\\ h_1^\pm\\
		h_2^\pm
	\end{array} \right), \quad \left( \begin{array}{c}
		\rho_2^\pm \\
		\chi^\pm
	\end{array} \right) = \left( \begin{array}{cc}
		- s_\theta & c_\theta \\
		c_\theta & s_\theta
	\end{array} \right) \left(\begin{array}{c}
		G_Y^\pm
		\\ h_3^\pm
	\end{array} \right),
	\label{EchargedH}
\end{eqnarray}
where $t_{\theta}=v_1/w$,
and 
\begin{align}\label{eq_fdepen}
	f&= \frac{c_{\beta } s_{\beta } \left(2 c_{\alpha }^2 m_{h^\pm_1}^2+2 s_{\alpha }^2 m_{h^\pm_2}^2-\tilde{\lambda}_{12} v^2\right)}{2 \omega ^2},
	%
	\;  f_{\eta}=\frac{\sqrt{2} c_{\alpha } s_{\alpha } (m_{h^\pm_2}^2-m_{h^\pm_1}^2)}{v},
	\crn \mu^2_{\sigma}&=\frac{1}{2} \left(2 c_{\alpha }^2 m_{h^\pm_2}^2-v^2 \left(c_{\beta }^2 \lambda _2^{\sigma }+s_{\beta }^2 \lambda _1^{\sigma }\right)+2 s_{\alpha
	}^2 m_{h^\pm_1}^2-\lambda _3^{\sigma } \omega ^2\right).
\end{align}
These result are consistent with Refs. \cite{Buras:2012dp, Hue:2017lak,Ninh:2005su} in the limits of $s_\alpha=0,\pm1$. The results given in Eqs. \eqref{EchargedH} and \eqref{eq_fdepen} obtained by  solving the following $3\times3$ squared mass matrix in the basis $(\eta^\pm,\; \rho_1^\pm,\; \sigma^\pm)$: 
\begin{align}
	\label{eq_33mc2}
	\mathcal{M}^2_{c}=\left(
	\begin{array}{ccc}
		\frac{f \omega ^2}{t_{\beta }}+\frac{1}{2} c_{\beta }^2 \tilde{\lambda }_{12} v^2 & f \omega ^2+\frac{1}{2} c_{\beta } \tilde{\lambda }_{12} s_{\beta } v^2 & \frac{c_{\beta } f_{\eta } v}{\sqrt{2}} \\
		f \omega ^2+\frac{1}{2} c_{\beta } \tilde{\lambda }_{12} s_{\beta } v^2 & f t_{\beta } \omega ^2+\frac{1}{2} \tilde{\lambda }_{12} s_{\beta }^2 v^2 & \frac{f_{\eta } s_{\beta } v}{\sqrt{2}} \\
		\frac{c_{\beta } f_{\eta } v}{\sqrt{2}} & \frac{f_{\eta } s_{\beta } v}{\sqrt{2}} & \frac{v^2}{2} \left(  c_{\beta }^2 \lambda _2^{\sigma } +s_{\beta }^2 \lambda_1^{\sigma }\right)  +\frac{\lambda _3^{\sigma }
			\omega ^2}{2}+\mu _{\sigma }^2 \\
	\end{array}
	\right).
\end{align}
We will find out that the Higgs masses $m_{h^\pm_{1,2}}$ and the mixing angle $\alpha$  are functions of the Higgs parameters in the Higgs potential.

The model contains five CP-odd neutral scalar components included in  the five neutral Higgs bosons  $\eta^0_1=(v_2 +R_1 + i  I_1)/\sqrt{2}$, $ \rho^0=(v_1 +R_2 + i  I_2)/\sqrt{2}$,  $ \chi^0_2=(\omega +R_3 + i  I_3)/\sqrt{2}$, $ \eta^0_2 =(R_4 + i  I_4)/\sqrt{2}$,  and $ \chi^0_1=(R_5 + i  I_5)/\sqrt{2}$.  Three of them are Goldstone bosons of  the neutral gauge bosons $Z,Z'$, and $X^0$. The two remaining are physical states with masses
\begin{align}
	m^2_{a_1}=\left(s_{\beta }^2 v^2+\omega ^2\right)  \left( f t^{-1}_{\beta} + \frac{1}{2}\tilde{\lambda}_{13} \right), \quad  m^2_{a_2}=f \left(\frac{\omega ^2}{c_{\beta }s_{\beta
	}}+c_{\beta } s_{\beta } v^2\right). 
\end{align} 
As a consequence, the parameter $f$ must satisfies $f>0$.  

Considering the CP-even scalars, 
there are  two sub-matrices $2\times2$ and $3\times 3$ for masses of these Higgs bosons in two bases $(\eta^0_2,\; \chi^0_1)$ and $(\eta^0_1,\; \rho^0_1, \chi^0_1)$, namely
\begin{align}
	M^2_{0,3}&= \left(
	\begin{array}{ccc}
		\frac{c_{\beta } f \omega ^2}{s_{\beta }}+2 s_{\beta }^2 \lambda _1 v^2 & c_{\beta } s_{\beta } \lambda _{12} v^2-\omega ^2 f & \omega  (s_{\beta } \lambda _{13}-c_{\beta } f) v \\
		c_{\beta } s_{\beta } \lambda _{12} v^2-\omega ^2 f & \frac{s_{\beta } f \omega ^2}{c_{\beta }}+2 c_{\beta }^2 \lambda _2 v^2 & \omega  (c_{\beta } \lambda _{23}-s_{\beta } f) v \\
		\omega  (s_{\beta } \lambda _{13}-c_{\beta } f) v & \omega  (c_{\beta } \lambda _{23}-s_{\beta } f) v & 2 \lambda _3 \omega ^2+c_{\beta } s_{\beta } f v^2 \\
	\end{array}
	\right),
	\crn M^2_{0,2}&= \left(
	\begin{array}{cc}
		\frac{1}{2} \omega ^2 \left(\tilde{\lambda }_{13}+\frac{2 c_{\beta } f}{s_{\beta }}\right) & \frac{1}{2} \omega  (\tilde{\lambda }_{13} s_{\beta }+2 c_{\beta } f) v \\
		\frac{1}{2} \omega  (\tilde{\lambda }_{13} s_{\beta }+2 c_{\beta } f) v & \frac{1}{2} s_{\beta } (\tilde{\lambda }_{13} s_{\beta }+2 c_{\beta } f) v^2 \\
	\end{array}
	\right). 
\end{align}
The matrix $M^2_{0,2}$ has one zero value and $m^2_{h_4}=\left(\frac{f}{t_{\beta }}+\frac{\tilde{\lambda }_{13}}{2}\right) \left(s_{\beta }^2 v^2+\omega ^2\right)$ corresponding to one Goldstone boson of $X^0$ and a heavy neutral Higgs boson $h^0_4$ with mass at the $SU(3)_L$ breaking scale.  On the other hand, we see that  Det$[ M^2_{0,3}] \neq0$ but Det$[\left.M^2_{0,3}]  \right|_{v=0}=0$, which implies that there is at least one Higgs boson mass at the electroweak scale that can be identified with the SM-like Higgs boson.   In particular, it can be proved that
\begin{align}\label{eq_Ch1}
	C^{h}_1	\left.M^2_{0,3}C^{hT}_1 \right|_{v=0}=\mathrm{diag}\left( 0,\; 2\lambda_3 w^2, fw^2/(s_\beta c_\beta)\right),\; C^{h}_1=\left(
	\begin{array}{ccc}
		s_{\beta } & c_{\beta } & 0 \\
		-c_{\beta } & s_{\beta } & 0 \\
		0 & 0 & 1 \\
	\end{array}
	\right),
\end{align}
and $C^{h}_1M^2_{0,3}C^{hT}_1\equiv\;M'^2_{0,3} $ satisfying:
\begin{align}
	\label{eq_Mp203}	
	\left( M'^2_{0,3}\right)_{11}&= 2 v^2 \left(c_{\beta }^4 \lambda _2+c_{\beta }^2 \lambda _{12} s_{\beta }^2+\lambda _1 s_{\beta }^4\right),
	\crn 
	\left( M'^2_{0,3}\right)_{22}&=2 c_{\beta }^2 s_{\beta }^2 v^2 (\lambda _1-\lambda _{12}+\lambda _2) +\frac{f \omega ^2}{c_{\beta } s_{\beta }},
	\crn 
	\left( M'^2_{0,3}\right)_{33}&= f c_{\beta } s_{\beta } v^2+2 \lambda _3 \omega ^2,
	\crn 
	\left( M'^2_{0,3}\right)_{12}&=\left( M'^2_{0,3}\right)_{21}=c_{\beta } s_{\beta } v^2 \left(s_{\beta }^2 (\lambda _{12}-2 \lambda _1)-c_{\beta }^2 (\lambda _{12}-2 \lambda _2)\right),
	\crn 
	\left( M'^2_{0,3}\right)_{13}&=\left( M'^2_{0,3}\right)_{31}= v \omega  \left(-2 f c_{\beta } s_{\beta }+c_{\beta }^2 \lambda _{23}+\lambda _{13} s_{\beta }^2\right),
	\crn 
	\left( M'^2_{0,3}\right)_{32}&=\left( M'^2_{0,3}\right)_{23} =v \omega  \left(f c_{\beta }^2-f s_{\beta }^2+c_{\beta } s_{\beta } (\lambda _{23}-\lambda _{13})\right).
\end{align}
Therefore, there is a unitary transformation $C^{h}_2$ with $\left( C^{h}_2\right)_{ij}\sim \mathcal{O}(v/w)$  ($i\neq j$) so that $C^{h}_2M'^2_{0,3}C^{hT}_2= \mathrm{diag}\left(m^2_{h^0_1},\; m^2_{h^0_2},\;m^2_{h^0_3} \right) $ and $m^2_{h^0_1}\sim \mathcal{O}(v^2)$  \cite{Okada:2016whh, Nguyen:2018rlb, Hung:2019jue}.   Hence  $h^0_1$ is identified with the SM-like Higgs boson found at the LHC, namely $h^0_1\equiv h$.  For simplicity, we will fix  $C^{h}_2=I_3$ in this work, and use the relations  $(\eta^0_1,\; \rho^0_1, \chi^0_1)= C^{hT}_1(h^0_1,\; h^0_2, h^0_3)$ in our numerical investigation, where only $\eta^0_1\sim R_1$ and $\rho^0_1\sim R_2$ give contributions to $h^0_1$, namely
\begin{equation}\label{eq_h01}
	R_1=s_{\beta} h^0_1 -c_{\beta} h^0_2,\; R_2= c_{\beta} h^0_1 +s_{\beta} h^0_2. 
\end{equation}
This assumption leads to a consequence that the $m_{h^0_1}$ is independent with the Higgs self-couplings relating with one-loop decays $h^0_1\to e_a e_b$, as it will be seen as follows:
\begin{equation}\label{eq_h01Hpm}
-\mathcal{L}_h=	V_{h}= \sum_{i,j=1}^3 - g_{hij} h^0_1h^+_{i}h^-_j+\dots.,
\end{equation}
where  non-zero $ g_{h^0_1ij}=g_{hji}$ are
\begin{align} \label{eq_h01hpimj}
	g_{h11}&=	-vc_{\alpha }^2 \left[  \left(2 c_{\beta }^2 s_{\beta }^2 (\lambda_1 -\lambda_{12} +\lambda _2)+\lambda _{12} +\tilde{\lambda }_{12}\right) +t_{\alpha }^2 \left(c_{\beta }^2  \lambda_2^{\sigma } +s_{\beta }^2 \lambda_1^{\sigma }\right) + \frac{2  s_{\alpha }^2 (m_{h^\pm_1}^2 -m_{h^\pm_2}^2)}{v^2}\right] ,
	\crn  g_{h22}&= -v c_{\alpha }^2\left[ t_{\alpha }^2 \left(2 c_{\beta }^2 s_{\beta }^2 (\lambda_1 -\lambda_{12} +\lambda _2)+\lambda _{12} +\tilde{\lambda }_{12}\right) + \left(c_{\beta }^2  \lambda_2^{\sigma } +s_{\beta }^2 \lambda_1^{\sigma }\right) - \frac{2  s_{\alpha }^2 (m_{h^\pm_1}^2 -m_{h^\pm_2}^2)}{v^2} \right]  ,
	\crn  g_{h12} &= -c_{\alpha } s_{\alpha } v\left[  2s_{\beta }^2 c_{\beta }^2 (\lambda_1 -\lambda_{12} +\lambda_2)   -s_{\beta }^2\lambda_1^{\sigma }  -c_{\beta }^2 \lambda_2^{\sigma}  +\lambda_{12} +\tilde{\lambda }_{12}  -\frac{ (c^2_{\alpha } -s^2_{\alpha })  (m_{h^\pm_1}^2-m_{h^\pm_2}^2)}{v^2} \right], 
	\crn  g_{h33}&= -v \left[c_{\beta }^2 \left(2 c_{\theta }^2 \lambda _2+s_{\theta }^2 (\lambda _{23}+\tilde{\lambda }_{23})\right)+s_{\beta }^2 \left(c_{\theta }^2 \lambda _{12}+\lambda _{13} s_{\theta }^2\right)   +c_{\beta } c_{\theta }^2  (2 f s_{\beta }+c_{\beta } \tilde{\lambda }_{23}) \right]  .
\end{align}

In the next section, we will derive all of the remaining couplings giving one-loop contributions of decays mentioned in this work.

\section{\label{sec_coupling} Couplings and analytic formulas}

\subsection{Decays $e_b\to e_a\gamma$ and $(g-2)_{e_a}$}
The couplings of charged gauge bosons giving one-loop contributions to LFV amplitudes are:
\begin{align}
	\label{eq_LVff}
	L_{V^\pm ff}&= \frac{g}{\sqrt{2}} \sum_{a=1}^3\sum_{i=1}^9\overline{n_i} \gamma^{\mu} P_L e_a \left[ U^{\nu*}_{ai}  W^{+}_{\mu} +U^{\nu*}_{(a+3)i}  Y^{+}_{\mu}\right] +\mathrm{h.c.},	
\end{align}
All the  calculation steps to derive theses  couplings were presented in Ref.~\cite{Nguyen:2018rlb}.  From now on, we always choose that $m_{e_b}>m_{e_a}$, equivalently $b>a=1,2,3$, to define the decays $e_b\rightarrow e_a \gamma$. 
One-loop form factors from charged gauge bosons are~\cite{Crivellin:2018qmi}:
\begin{align}
	\label{eq_cabVR}
	c_{(ab)R}(W)&= \frac{e g^2 }{32\pi^2 m_W^2} \sum_{i=1}^9 U^{\nu}_{ai}U^{\nu*}_{bi} \tilde{F}_{V}\left( x_{W,i}\right), \crn 
	c_{(ab)R}(Y)&= \frac{e g^2 }{32\pi^2 m_Y^2} \sum_{i=1}^9  U^{\nu}_{(a+3)i}U^{\nu*}_{(b+3)i} \tilde{F}_{V}\left( x_{Y,i}\right),
\end{align}
where $x_{v,i}=m^2_{n_i}/m^2_v$; $v=W,Y$; 
\begin{equation}\label{eq_FFVV}
	\tilde{F}_{V}(x)= -\frac{10 -43 x +78 x^2 -49 x^3 +4 x^4 +18 x^3\ln(x)}{24(x-1)^4};
\end{equation}
$e=\sqrt{4\pi \alpha_{\mathrm{em}}}$ being the electromagnetic coupling constant;  and $g=e/s_W$. 

The Yukawa couplings of charged Higgs bosons  with leptons  are defined by
\begin{align}
	\label{eq_LHee}
	\mathcal{L}^{\ell n h^\pm}=- \frac{g}{\sqrt{2}m_W} \sum_{k=1}^3 \sum_{a=1}^3\sum_{i=1}^9 h_k^+\overline{n_i} \left(\lambda^{L,k}_{ai}P_L+\lambda^{R,k}_{ai}P_R\right)e_a   +\mathrm{h.c.},
\end{align} 
where  
\begin{align}
	\lambda^{R,1}_{ai}&=m_{e_a}c_{\alpha} t_{\beta}U^{\nu*}_{ai} -\sum_{c=1}^3 \frac{vY^{\sigma}_{ca}s_{\alpha}}{\sqrt{2}} U^{\nu*}_{(c+6)i},  	\quad  \lambda^{L,1}_{ai} = c_{\alpha}s_{\beta}z_0e^{i\alpha_{23}}\sum_{c=1}^3(\tilde{m}_D)_{ac}U^{\nu}_{(c+3)i},
	\crn  	\lambda^{R,2}_{ai}&=m_{e_a}s_{\alpha}t_{\beta}U^{\nu*}_{ai} +\sum_{c=1}^3 \frac{vY^{\sigma}_{ca}c_{\alpha}}{\sqrt{2}} U^{\nu*}_{(c+6)i},  \quad 
	\lambda^{L,2}_{ai} =  s_{\alpha}s_{\beta}z_0e^{i\alpha_{23}}\sum_{c=1}^3(\tilde{m}_D)_{ac}U^{\nu}_{(c+3)i},	
	\crn 	\lambda^{R,3}_{ai}&=\frac{m_{e_a}c_{\theta}U^{\nu*}_{(a+3)i}}{c_{\beta}}, \quad \lambda^{L,3}_{ai} = c_{\theta}z_0 \sum_{c=1}^3 \left[ -e^{i\alpha_{23}} (\tilde{m}_D)_{ac}U^{\nu}_{ci} + t^2_{\theta}(\tilde{M}^T_R)_{ac}U^{\nu}_{(c+6)i}\right] .
	\label{eq_lambdaLR}
\end{align}

The interactions given in Eqs. \eqref{eq_LVff} and \eqref{eq_LHee} also give tree and  loop contributions to the lepton flavor conserved  decay $\mu^-\to e^-\overline{\nu}_e \nu_{\mu} $.   Regarding the gauge couplings given in Eq. \eqref{eq_LVff}, the couplings of $Y^\pm$ with active neutrinos are zeros  because $U^{\nu}_{(c+3)1}=U^{\nu}_{(c+3)2}=0$, the difference of the  couplings of  $W$ with active neutrinos and charged leptons between the SM and the 331ISS model under consideration is $|\frac{1}{2}(R_2R^+_2U)_{ab}|\ll1$. Regarding the Higgs boson contributions, only $\lambda^{L,3}_{ai}$ may give large contributions to the decay amplitude  $\mu^-\to e^-\overline{\nu}_e \nu_{\mu} $, because the remaining couplings are always proportional to $gm_{\mu} t_{\beta}/m_W \ll1$ or $U^{\nu}_{(c+3)2}U^{\nu}_{(c+3)1}=0$. Assuming $t_{\theta}=0$ for very large $SU(3)_L$ scale $w\gg v$, we have a crude approximation that $|\lambda^{L,3}_{ai}|\leq z_0$. The  large values of  $|\lambda^{L,3}|$  appear  because $h^\pm_3\simeq \rho^\pm_2$, which  has couplings with active neutrinos  $\overline{e_a}\left(\nu_{bL}\right)^c\rho^-_2\sim h^{\nu}_{ab}\sim (\tilde{m}_D)_{ab}$ derived from the second term in Lagrangian \eqref{eq_Lylepton}.  Based on the well-known formulas of the partial decay width $\Gamma(\mu\to 3e)$ at tree level given in the Zee-Babu model \cite{Nebot:2007bc}, the coupling $\lambda^L$ leads to a deviation of the decay width of  the  decay $\mu^-\to e^-\overline{\nu}_e \nu_{\mu} $ between the 331ISS model and the SM as follows: 
\begin{align}
	\label{eq_deGamuenmue}
	|\delta \Gamma^{\mathrm{331ISS}}(\mu^-\to e^-\overline{\nu}_e \nu_{\mu})| &\equiv \left| \frac{\Gamma^{\mathrm{331ISS}}(\mu^-\to e^-\overline{\nu}_e \nu_{\mu})}{\Gamma^{\mathrm{SM}}(\mu^-\to e^-\overline{\nu}_e \nu_{\mu})} -1\right|	 
	\crn& \simeq \left[ \frac{|\lambda^{L,3}_{ai}|^2}{4m^2_{h^\pm_3}}\right]^2= \left[\frac{|z_0^2|}{4m^2_{h^\pm_3}}\right]^2 \leq 10^{-6}.
\end{align}
The  constraint is derived from the mean life time of muon \cite{ParticleDataGroup:2020ssz}. The derivation of the formula \eqref{eq_deGamuenmue} is summarized as follows.  The total amplitude is $i\mathcal{M}= i\mathcal{M}_W +i\mathcal{M}_{h^\pm}$, where $\mathcal{M}_W$ and $\mathcal{M}_{h^\pm}$ are   the contributions from $W$  and charged Higgs bosons, respectively. In the low energy limit we have  
\begin{align*}
	\mathcal{M}_W&\simeq \mathcal{M}^{\mathrm{SM}}\sim \frac{g^2}{2m_W^2} [\overline{u}_{\nu_{\mu}}\gamma^{\mu}P_Lu_{\mu}][\overline{u}_{e} \gamma_{\mu}P_Lv_{\nu_e}],
	\crn \mathcal{M}_{h^\pm}&\sim \frac{g^2}{2m_W^2 m^2_{h^\pm}} \times [\overline{u}_{\nu_{\mu}}\left( \lambda^LP_L+\lambda^RP_R\right)u_{\mu}][\overline{u}_{e} \left( \lambda^{L*}P_R+\lambda^{R*}P_L\right)v_{\nu_e}]. 
\end{align*} 
Now it can be proved that $|\mathcal{M}|^2=|\mathcal{M}_W|^2 +|\mathcal{M}_{h^\pm}|^2$ because $\mathcal{M}_W^*\mathcal{M}_{h^\pm}$ has an  odd number of the gamma matrices in the trace and $m_e,m_{\nu_\mu},m_{\nu_e}\simeq0$, leading to $\mathcal{M}_W^*\mathcal{M}_{h^\pm}=0$. 

In the numerical investigation, we will choose $m_{h^\pm_3}\ge z_0\times 10\sqrt{5}$ to accommodate the constraint \eqref{eq_deGamuenmue}. Now we can assume the approximation that $\Gamma^{\mathrm{331ISS}}(\mu^-\to e^-\overline{\nu}_e \nu_{\mu})\simeq \Gamma^{\mathrm{SM}}(\mu^-\to e^-\overline{\nu}_e \nu_{\mu})$. This  approximation   for calculating the cLFV decay rates  is consistent with many works  published recently \cite{Enomoto:2019mzl, Camara:2020efq}.

The one-loop form factors are ~\cite{Crivellin:2018qmi}:
\begin{align}
	\label{eq_cabHR}
	c_{(ab)R}(h^\pm_k)&= \frac{e g^2}{32\pi^2m^2_W m_{e_b}m^2_{h^\pm_k}} \sum_{i=1}^9\left[ \lambda^{L,k*}_{ai} \lambda^{R,k}_{bi} m_{n_i}F_{H}\left(x_{k,i}  \right) 
	\right. \crn& \quad \left.
	+  \left( m_{e_b}\lambda^{L,k*}_{ai} \lambda^{L,k}_{bi} +m_{e_a} \lambda^{R,k*}_{ai} \lambda^{R,k}_{bi}\right) \tilde{F}_{H}\left(x_{k,i} \right)\right], 
 \end{align}
where  $b\ge a$, $x_{k,i}=m^2_{n_i}/m^2_{h^\pm_k}$, and the one-loop functions $F_{H}(x)$ and  $\tilde{F}_{H}(x)$ are  
\begin{equation}  \label{eq_FH}
	F_{H}(x)= -\frac{1 -x^2 + 2x\ln(x)}{4(x-1)^3}, \quad \widetilde{F}_{H}(x)= -\frac{-1 +6 x -3 x^2  -2 x^3  + 6 x^2\ln(x)}{24(x-1)^4}.
\end{equation}
The total one-loop contributions to the cLFV amplitude $e_b\to e_a \gamma$ and $\Delta a^{331\mathrm{ISS}}_{e_a}$ are  
\begin{align} \label{eq_cabR}
	c_{(ab)R}&=\sum_{x=W,Y}c_{(ab)R} (x)+ \sum_{k=1}^3c_{(ab)R}(h_k^\pm), \crn
	c_{(ba)R}&=\left( c_{(ab)R}\left[a\leftrightarrow b\right] \right)\times \frac{m_{e_a}}{m_{e_b}}.
\end{align}
 The second line of Eq. \eqref{eq_cabR} is derived from the  equality that $c_{(ba)R}(x)= \left(c_{(ab)R}(x) \left[ b\leftrightarrow a\right]\right) \times (m_{e_a}/m_{e_b})$ for all $x=W,Y,h^{\pm}_{1,2,3}$.  The formulas for the  contributions  to $a_{e_a}$  are:
\begin{align}
	\label{eq_aea}
	a_{e_a}& =-\frac{4 m^2_{e_a}}{e} \mathrm{Re}[c_{(aa)R}]=  -\frac{4 m^2_{e_a}}{2\pi^2 v^2} \mathrm{Re}[c'_{(aa)R}], \quad c'_{(ab)R}=c_{(ab)R} \times \left( \frac{eg^2}{32\pi^2 m^2_W}\right)^{-1}. 
\end{align}
One-loop contributions from heavy neutral Higgs bosons are very suppressed, hence they are ignored  here.   The deviation of $a_{e_a}$ between predictions by  the two models 331ISS and SM are 
\begin{align}
	\label{eq_defDamu331}
	\Delta a_{e_a}= \Delta a^{\mathrm{331ISS}}_{e_a}&\equiv  a_{e_a} - a^{\mathrm{SM}}_{e_a}(W),   
\end{align}
where $a^{\mathrm{SM}}_{\mu}(W)= 5g^2m^2_{\mu}/(96\pi^2 m^2_W) $ is the SM's prediction \cite{Jegerlehner:2009ry}.  In this work, $\Delta a_{e_a}$ will be considered as new physics (NP) predicted by the 331ISS, used to compare with experimental data in  numerical investigations.

The branching ratios   of the cLFV  processes  are 
\cite{Crivellin:2018qmi}
\begin{align}
	\label{eq_Gaebaga}
	\mathrm{Br}(e_b\rightarrow e_a\gamma) &\simeq  \frac{6\alpha_{em}}{ \pi} \left( \left|c'_{(ab)R}\right|^2 +\left|c'_{(ba)R}\right|^2\right) \mathrm{Br}(e_b\rightarrow e_a\overline{\nu_a}\nu_b),
\end{align}
where $G_F=1/(\sqrt{2}v^2)$,  consistent with  previous results \cite{Hue:2017lak, Nguyen:2018rlb} for 3-3-1 models.  

The formulas of $U^{\nu}$ given in Eq. \eqref{eq_Unu1} results in  approximate expressions  of $c_{(ab)R}$ and $c_{(ba)R}$ with $b\geq a$ as follows:
\begin{align}
	c'_{(ab)R}(W)&= -\frac{5}{12} \left[\delta_{ab} -(\tilde{m}^{\dagger}_DV_R\hat{k}^{-2}\tilde{m}_D)_{ab}\right] + \sum_{e=1}^3 (\tilde{m}_D^{\dagger}V_R^*\hat{k}^{-1})_{ae} (\tilde{m}_D^{T}V_R\hat{k}^{-1})_{be} F_V(x'_{W,e}),
	\crn 	c'_{(ab)R}(Y)&=  \frac{m_W^2}{m^2_Y} \sum_{e=1}^3 (V_R)^*_{ae} V_R)_{be} F_V(x'_{Y,e}),
	\crn 	c'_{(ab)R}(h^\pm_1)&= 
	\frac{z_0^2}{m^2_{h^\pm_1}}  \sum_{e=1}^3(\tilde{m}^{*}_DV^*_R)_{ae}   \left\{ c^2_{\alpha} s^2_{\beta}   (\tilde{m}_D^TV_R \hat{k}^{-1})_{be}  -\frac{vs_{2\alpha}  s_{2\beta}}{4m_b} \left[Y^{\sigma T}V_R \right]_{be}\right\} \hat{k}_e F_H(x'_{e,1}) 
	\crn &+ \frac{c^2_{\alpha} s^2_{\beta}z_0^2}{m^2_{h^\pm_1}}\sum_{e=1}^3  (\tilde{m}_DV_R)^{*}_{ae} (\tilde{m}_DV_R)_{be} \tilde{F}_H(x'_{e,1})
	\crn&+\frac{1}{24}\left\{\frac{m^2_{e_a} c^2_{\alpha}t^2_{\beta}}{m^2_{h^\pm_1}}\delta_{ab}   
	 + \frac{m_{e_a}vs_{\alpha}c_{\alpha}t_{\beta}}{\sqrt{2}m^2_{h^\pm_1}} \left[ \frac{m_{e_a}}{m_{e_b}} \left(R_2 Y^\sigma\right)_{ab} + \left(Y^{\sigma\dagger}R^{\dagger}_2 \right)_{ab}\right]\right\}
	 \crn&+ \sum_{e=1}^3 \tilde{F}_H(x'_{e,1}) c^2_{\alpha} \left\{ \frac{m^2_{e_a}t^2_{\beta}}{m^2_{h^\pm_1}} \left[ (R_2V_L)_{ae}(R_2V_L)^*_{be}\right]
	\right.
	+ \frac{v^2t^2_{\alpha}m_{e_a}}{m_{e_b}m^2_{h^\pm_1}} \left[ (Y^{\sigma T}V_R)_{ae} (Y^{\sigma T}V_R)^*_{be}  \right] 
	\crn &-\left. \frac{m_{e_a} vs_{\alpha}c_{\alpha}t_{\beta}}{m^2_{h^\pm_1}}\left[ \frac{m_{e_a}}{m_{e_b}} (R_2V_L)_{ae}(Y^{\sigma T}V_R)^*_{be} + (Y^{\sigma T}V_R)_{ae} (R_2V_L)^*_{be}\right] \right\}, 	\label{eq_app}
	\\ 	c'_{(ab)R}(h^\pm_2)&=c'_{(ab)R}(h^\pm_1) \left[m_{h^\pm_1} \to m_{h^\pm_2}, c_{\alpha} \to s_{\alpha}, s_{\alpha} \to -c_{\alpha}\right],
	\crn 	c'_{(ab)R}(h^\pm_3)&= \frac{z_0^2}{m^2_{h^\pm_3}} \left\{ \sum_{e=1}^3 \left[ \left( \tilde{m}^*_D \tilde{m}_D V_R \hat{k}^{-1} \right)_{ae} \left(V^*_R\right)_{be} k_e F_H(x'_{e,3})\right]   \right.  
	%
	- \frac{1}{24}\left( \tilde{m}^*_D \tilde{m}_D 
	\right)_{ab} 
	\crn& \left. +\sum_{e=1}^3 \left[ \left( \tilde{m}_D \tilde{m}^*_D V_R \hat{k}^{-1} \right)_{ae}\left( \tilde{m}^*_D \tilde{m}_D V^*_R \hat{k}^{-1} \right)_{be}  + \frac{m_{e_a}^2}{z_0^2c^2_{\beta}}  \left(V_R^*\right)_{be}\left(V_R\right)_{ae} \right]  \tilde{F}_H(x'_{e,3})
 \right\},\nn 
\end{align}
where  equalities  in Eq. \eqref{eq_R2VL} were used. In addition, we ignore the minor contributions proportional to $R^\dagger_2 R_2$, and $R_2 R_2^\dagger$.   Because only two terms relating to $R_2Y^\sigma$ and $R_2^{\dagger}R_2$ depend on $V_L$, but   give small one-loop contributions to $\Delta a_{e_a}$,  we fix  $V_L=I_3$ without loss of generality. 

 The above expressions of $c_{(ab)R}$ and $c_{(ba)R}$ given in Eq. \eqref{eq_app} give some interesting properties.  First, all terms are  proportional to $1/m^{2}_{h^\pm_k}$, hence large $|\Delta a_{e_a}|$  corresponding to large  $|c_{(aa)R}|$ will prefer small $m^2_{h^{\pm}_k}$. In contrast, experimental constraints  on cLFV decay rates require small $|c_{(ab)R}|$ and $|c_{(ba)R}|$, hence $m^{2}_{h^\pm_k}$  should be large. It is easy to get small Br$(e_b\to e_a\gamma)$ with enough large $m_{h^\pm_k}$, but difficult  to get large $|\Delta a_{e_a}|$. Previously numerical investigation has showed another situation \cite{Hue:2021xap}, where small $m^2_{h^\pm_k}$ are needed for large $\Delta a_{\mu}$, and the destructive correlations between particular terms in $c_{(ab)R}$ and $c_{(ba)R}$ must appear to  result in small Br$(e_b\to e_a \gamma)$. The structure of the  mass Dirac matrix $\tilde{m}_D$ strongly affects these destructive correlations. As we will see, the antisymmetric property of  $\tilde{m}_D$ and the neutrino oscillation data fix a certain form of $\tilde{m}_D$, namely the fixed values considered in this work are   $(\tilde{m}_D)_{32}=-(\tilde{m}_D)_{23}=1$, $-(\tilde{m}_D)_{12}=(\tilde{m}_D)_{21} \simeq 0.613$, and $-(\tilde{m}_D)_{13}= (\tilde{m}_D)_{31} \simeq 0.357$, and $(\tilde{m}_D)_{11}=(\tilde{m}_D)_{22}=(\tilde{m}_D)_{33}=0$.  They do not support large absolute values  of the diagonal entries relating to $c_{(aa)R}$. Therefore, the simple case of $V_R=I_3$, degenerate values of heavy neutrino masses $\hat{k}_{11}=\hat{k}_{22}=\hat{k}_{33}$, and $Y^{\sigma}=\mathcal{O}_{3\times3}$ will give $c_{(ab)R}\sim \tilde{m}_D \tilde{m}^*_D $. As a result, constraints on cLFV decays always exclude the regions of parameter space predicting large $(g-2)_{e, \mu}$. This conclusion is completely consistent with the numerical results reported in Ref. \cite{Hue:2021xap}.  In addition, the presence of  $\sigma^\pm$ and non-zero Yukawa coupling matrix $Y^{\sigma}$ is necessary to explain $1\sigma$ range of $(g-2)_{\mu}$ obtained by experiment. Additionally, the formulas given in  Eq. \eqref{eq_app} explain explicitly that  large $\Delta a_{\mu}$ also needs large $z_0$. And,  large $t_{\beta}$ and non-zero $Y^{\sigma}$ support more strong  destructive correlations  to guarantee that $(e_b\to e_a)$ satisfies the current constraints. 
 
 Finally,  We emphasize that  the $(g-2)_e$ data and LFVH decays   were not discussed previously for  the 331ISS model. Our numerical investigation showed that large $(g-2)_e$ requires nonzero values of $s_{\alpha}$, which was not considered in Ref. \cite{Hue:2021xap}. In addition, large values of $Y^{\sigma}_{22,33,23,32}$ should be investigated carefully because they may result in too large Br$(h\to \tau\mu)$ that may be excluded by the experimental constraints. 

\subsection{Decays $h^0_1\to e_ae_b$} 
The Yukawa couplings  $h^0_1ff$  is , namely
\begin{align}
	\mathcal{L}^Y_{h^0_1ff}=-\frac{g}{2m_W} h^0_1 \left[ \frac{1}{2}\sum_{i,j=1}^9 \overline{n_i} \left( \lambda^0_{ij} P_L+\lambda^{0*}_{ij} P_R \right)n_j  +m_{e_a} \overline{ e_a} e_a \right],
\end{align}
where
\begin{equation}\label{eq_la0ij}
	\lambda^0_{ij}= 
	\sum_{c=1}^3\left(U^{\nu}_{ci}U^{\nu*}_{cj}m_{n_i}+U^{\nu*}_{ci}U^{\nu}_{cj}m_{n_j}\right),
\end{equation}  
is    a symmetric coefficient $\lambda^0_{ij}=\lambda^0_{ji}$  corresponding to the Feynman rules given in Ref. \cite{Dreiner:2008tw}.  All of the Feynman rules for couplings involved in LFV processes at one-loop level  are listed in Table~\ref{numbers}, 
\begin{table}[h]
	\begin{tabular}{|c|c|}
		\hline
		Vertex & Coupling \\
		\hline
		$ h^0_1 \overline {e_a}e_a$ & $-\frac{i gm_{e_a}}{2m_W}$\\
		\hline
		$ h^0_1 \overline {n_i}n_j$ & $- \frac{ig}{2m_W} \left( \lambda^0_{ij} P_L+\lambda^{0*}_{ij} P_R \right)$\\
		\hline
		$  h_k^+\overline{n_i} e_b$, 	$h_k^-\overline {e_a} n_i $ & $\frac{-ig}{\sqrt{2}m_W} \left( \lambda^{L,k}_{bi} P_L + \lambda^{R,k}_{bi} P_R\right)$,  $\frac{-ig}{\sqrt{2}m_W} \left( \lambda^{L,k*}_{ai} P_R + \lambda^{R,k*}_{ai} P_L\right)$\\
		\hline
		$ W_\mu^+ \overline{ n_i} e_b$,  	$   W_\mu^-\overline {e_a}n_i$ & $\frac{ig}{\sqrt{2}} U^{\nu*}_{ai}\gamma^{\mu}P_L$,   $\frac{ig}{\sqrt{2}} U^{\nu}_{ai}\gamma^{\mu} P_L$ \\
		\hline
		$Y_\mu^+ \overline{n_i}  e_b$, 	$Y_\mu^-  \overline {e_a} n_i $&   $\frac{ig}{\sqrt{2}} U^{\nu*}_{(a+3)i}\gamma^{\mu}P_L$,   $\frac{ig}{\sqrt{2}} U^{\nu}_{(a+3)i}\gamma^{\mu}P_L$\\
		\hline
		$h_3^+h^0_1Y_{\mu}^-$, $h_3^-Y^+_{\mu}h^0_1$ &  $\frac{i}{2}gc_{\beta}c_{\theta}(p_+ -p_0)^{\mu}$, $-\frac{i}{2}gc_{\beta}c_{\theta}(p_- -p_0)^{\mu}$\\
		\hline
		$h^0_1 W^+_{\mu }W^-_{\nu}$ & $igm_Wg^{\mu \nu}$ \\
		\hline
		$h^0_1 Y^+_{\mu }Y^-_{\nu}$ & $igc_{\beta}s_{\theta} m_Yg^{\mu \nu}$\\
		\hline
	\end{tabular}
	\caption{ Feynman rules for one-loop contributions to $(g-2)$ anomalies, $e_b \to e_a \gamma$, and  $h^0_1\to e_ae_b$ in the unitary gauge. $p_0$ and $p_{\pm}$   are the incoming momenta  of $h^0_1$ and $h_3^{\pm}$, respectively.    \label{numbers}}
\end{table}
where we used $s_{\theta}=gv_1/(2m_Y)$. We  focus on the limit of tiny $t_{\theta}\simeq s_{\theta}=0$, and the suppressed  deviation  of the  SM-like Higgs mixing mentioned previously \cite{Okada:2016whh, Nguyen:2018rlb}. Namely, they will be fixed to be zeros in the numerical calculations. 

\begin{figure}[h]
	\centering
	\includegraphics[width=14cm]{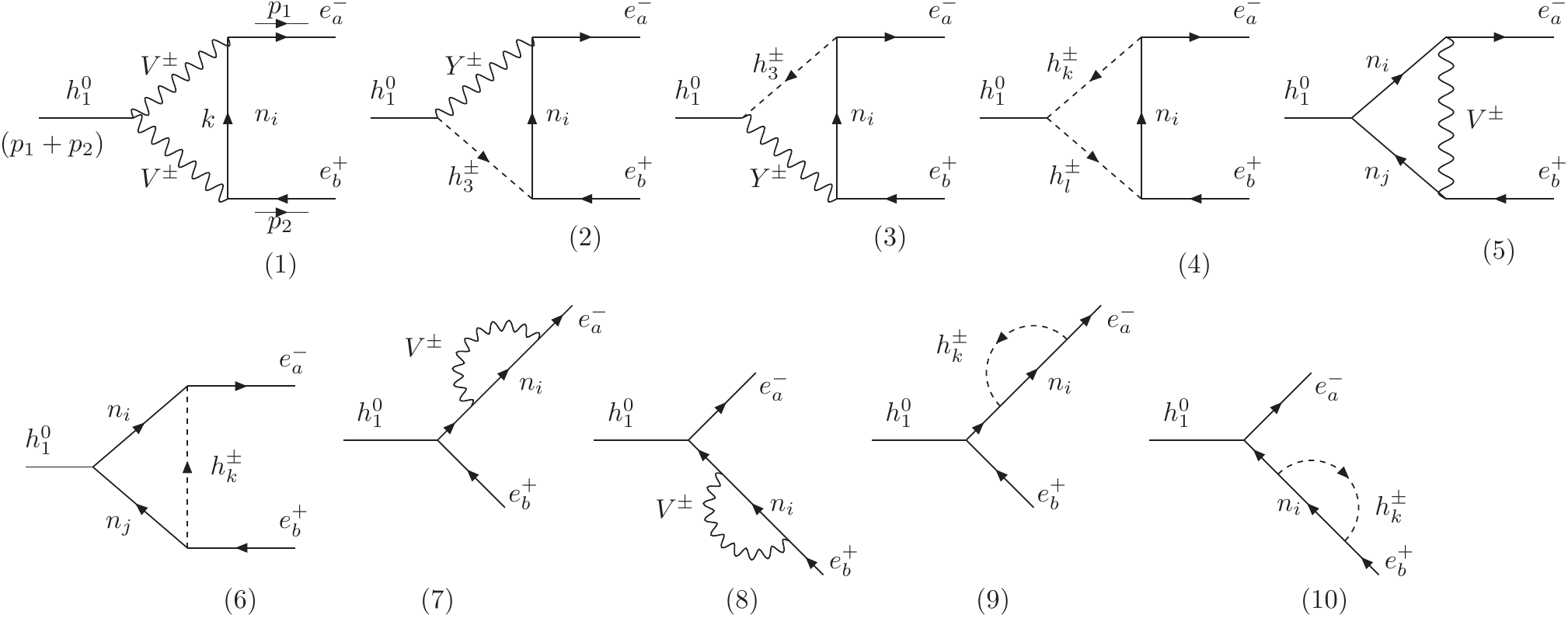}\\
	\caption{One-loop Feynman diagrams contributing to the decay $h^0_1\rightarrow e_a e_b$ in the unitary gauge.  Here $V^\pm=W^{\pm},\,Y^{\pm}$; $k,l=1,2,3$. }\label{hlilj1}
\end{figure}

The effective Lagrangian and partial decay width of the decay  $h^0_1\rightarrow e_a^{\pm}e_b^{\mp}$  are 
\begin{align}
\mathcal{L}^{\mathrm{LFVH}}&= h^0_1 \left(\Delta_{(ab)L} \overline{e_a}P_L e_b +\Delta_{(ab)R} \overline{e_a}P_R e_b\right) + \mathrm{H.c.}, 
\crn 	\Gamma (h_1^0\rightarrow e_ae_b)&= \Gamma (h_1^0\rightarrow e_a^{-} e_b^{+})+\Gamma (h^0_1\rightarrow e_a^{+} e_b^{-})
=  \fr{ m_{h^0_1} }{8\pi }\left(\vert \Delta_{(ab)L}\vert^2+\vert \Delta_{(ab)R}\vert^2\right), \label{eq_LFVwidth}	
\end{align}
where the scalar factors $\Delta_{(ab)L,R}$  are  loop contributions in this work. In the unitary gauge, the one-loop Feynman diagrams contributing to $\Delta_{(ab)L,R}$ are shown in Fig. \ref{hlilj1}.
The valid condition  $m_{h_1^0}\gg m_{a,b}$ was used in Eq. \eqref{eq_LFVwidth}, where $m_{a,b}$ are the lepton masses satisfying   $p^2_{1,2}=m_{a,b}^2$  and $ p_{h_1^0}^2 \equiv( p_1+p_2)^2=m^2_{h_1^0}$. The branching ratio of a LFVH decays  is  Br$(h^0_1\rightarrow e_ae_b)= \Gamma (h_1^0\rightarrow e_ae_b)/\Gamma^{\mathrm{total}}_{h_1^0},$ where $\Gamma^{\mathrm{total}}_{h_1^0}\simeq 4.1\times 10^{-3}$ GeV \cite{Denner:2011mq,  ParticleDataGroup:2020ssz}.  The $\Delta_{(ab)L,R}$ can be written as
\begin{equation}\Delta_{(ab)L,R} =\sum_{i=1,5,7,8} \Delta^{(i)W}_{(ab)L,R} + \sum^{10}_{i=1} \Delta^{(i)Y}_{(ab)L,R},  \label{deLR}
\end{equation}
where the analytic forms of  $\Delta^{(i)W}_{(ab)L,R}$ and $\Delta^{(i)Y}_{(ab)L,R}$ are  shown in the Appendix.  There are numbers of  tiny one-loop contributions, which we will ignore  in the numerical calculations.  They are calculated using the unitary gauge with the same techniques given in Refs. \cite{Thao:2017qtn, Nguyen:2018rlb}.   The contributions from diagrams (2), (3), and (5) with $Y^\pm$ exchanges have suppressed factors $c_{\beta}m^3_W/m^3_Y$.  The  one-loop contributions from diagram (6) are suppressed with heavy singly charged Higgs bosons, which we checked consistently with the result mentioned in Refs. \cite{Nguyen:2018rlb, Jurciukonis:2021izn}. 

\section{\label{sec_numerical} Numerical discussion}

In this work, we will use  the neutrino oscillation data given in Refs. 
\cite{ParticleDataGroup:2020ssz, T2K:2019bcf}. The standard form of the lepton mixing matrix $U_{\mathrm{PMNS}}$ is the function   of three angles $\theta_{ij}$, one Dirac phase $\delta$ and two Majorana phases $\alpha_{1}$ and $\alpha_2$~\cite{ParticleDataGroup:2018ovx}, namely
\begin{align}
	U^{\mathrm{PDG}}_{\mathrm{PMNS}}&=f(s_{12},s_{13},s_{23},\delta)\times  \;\mathrm{diag}\left(1, e^{i\alpha_{1}},\,e^{i\alpha_{2}}\right),
\crn f(s_{12},s_{13},s_{23},\delta)&\equiv  \begin{pmatrix}
		1	& 0 &0  \\
		0	&c_{23}  &s_{23}  \\
		0&  	-s_{23}& c_{23}
	\end{pmatrix}\,\begin{pmatrix}
		c_{13}	& 0 &s_{13}e^{-i\delta}  \\
		0	&1  &0  \\
		-s_{13}e^{i\delta}&  0& c_{13}
	\end{pmatrix}\,\begin{pmatrix}
		c_{12}	& s_{12} &0  \\
		-s_{12}	&c_{12}  &0  \\
		0& 0 	&1
	\end{pmatrix},  \label{eq_UnuPDG}
\end{align}
where $s_{ij}\equiv\sin\theta_{ij}$, $c_{ij}\equiv\cos\theta_{ij}=\sqrt{1-s^2_{ij}}$, $i,j=1,2,3$ ($i<j$), $0\le \theta_{ij}<90\; [\mathrm{Deg.}]$ and $0<\delta\le 360\;[\mathrm{Deg.}]$. The Majorana phases are chosen in the range $-180\le\alpha_i\le 180$ [Deg.].  For numerical investigation,  we  choose a benchmark corresponding to the normal order  of the neutrino oscillation data as the input to fix $\tilde{m}_D$ that $s^2_{12}=0.32$, $s^2_{23}= 0.547$, $s^2_{13}= 0.0216$,  $\Delta m^2_{21}=7.55\times 10^{-5} [\mathrm{eV}^2],$ $ \Delta m^2_{32}=2.424\times 10^{-3} [\mathrm{eV}^2]$, $\delta= 180 \;[\mathrm{Deg}] $, and $\alpha_1=\alpha_2=0$. Consequently, the reduced Dirac mass matrix $\tilde{m}_D$ is  fixed as 
\begin{equation}\label{eq_tmD}
\tilde{m}_D=\begin{pmatrix}
0	&0.613  & 0.357 \\
-0.613	& 0 & 1 \\
-0.357	&-1  &0 
\end{pmatrix}.
\end{equation}
The best-fit point for the normal (inverted) order is $\delta= -1.89^{+0.7}_{-0.58} (-1.38^{+0.48}_{-0.54}) \neq 180$ deg. \cite{T2K:2019bcf}, which rules out the value $180$ deg. at 95\% confidence level.  But it is  still allowed  in $3\sigma$ range. The other quantities corresponding to the best-fit point are $s^2_{23}=0.53$, $ \Delta m^2_{21}=7.53  \times 10^{-5}\;[\mathrm{eV^2}]$,  $\Delta m^2_{32}=2.45  \times 10^{-3}\; [\mathrm{eV^2}]$, leading  to a new $\tilde{m}_D$ with $(\tilde{m}_D)_{12}=0.546 e^{0.18 i}$ and  $(\tilde{m}_D)_{13}=0.453 e^{-0.23 i}$. The existence of  the non-zero CP violation  $\delta \neq$ 180 deg. will  lead to the complex values of the two entries of $\tilde{m}_{D}$ instead of the real ones given in Eq. (52).  These imagine parts   result in non-zero values of Im[$c_{(ab)R}$], which is enough to give large Br$(\mu \to e\gamma)  >4.2\times 10^{-13}$  in many regions of the parameter space, even when Re[$c_{(ab)R}$]=0. Therefore, many very complicated relations between  parameters must be satisfied to guarantee that all Im and Re parts contributing to these cLFV decays satisfy the experimental constraints.  In this work,    the  limit that $\delta= 180 \;[\mathrm{Deg}] $ is fixed for simplicity.

The mixing matrix $V_R$ is parameterized using the same formulas given in Eq. \eqref{eq_UnuPDG},   $V_{R}= f(s^r_{12},s^r_{13},s^r_{23},0)$  with  $|s^r_{ij}|\leq1$. The remaining free parameters are scanned in the following ranges:
\begin{align} \label{eq_scanRange}
\hat{k}_{1,2,3}&\geq5,\; 600 \;\; [\mathrm{GeV}]\leq m_{h^\pm_{1,2}}\leq 1500\; [\mathrm{GeV}],\; s_{\alpha}|\leq1, \; \mathrm{max}[|Y^{\sigma}_{ij}|]\leq 1.5,
\crn t_{\beta}&\in [30,70],\; 400\; [\mathrm{GeV}]\leq z\leq 1200 \; [\mathrm{GeV}],
\end{align}
and $m_{h^\pm_{3}}\;=40$ TeV, so that the decay width of $\mu^- \to e^-\overline{\nu_e}\nu_{\mu}$ is consistent with that predicted by the SM. In addition, the collected points satisfy that  max$|(R_2R^{\dagger}_2)_{ab}|<10^{-3}$ with all $a,b=1,2,3$.  This constraints also satisfies many other  recent experimental results such as  electroweak precision tests,  cLFV decays~\cite{Fernandez-Martinez:2016lgt,  Agostinho:2017wfs, Coutinho:2019aiy,  Manzari:2020eum}. The experimental parameters are $G_F=1.663787\times 10^{-5}\;[\mathrm{GeV}^{-2}]$, $g=0.652$, $\alpha_{em}=e^2/(4\pi)=1/137$, $s^2_W=0.231$, $m_e=5\times 10^{-4}$ [GeV], $m_{\mu}=0.105$ [GeV], $m_{\tau}=1.776$ [GeV], and $m_W=80.385$ [GeV], Br$(\mu\to e\overline{\nu_e}\nu_{\mu})\simeq 1$,   Br$(\tau\to e\overline{\nu_e}\nu_{\tau})\simeq 0.1782$, and  Br$(\tau\to \mu\overline{\nu_{\mu}}\nu_{\tau})\simeq 0.1739$.  We note that the upper bounds  of  $m_{h^\pm_{1,2}}$ and $t_{\beta}$ based on the previous work to accommodate large values of   $\Delta a_{\mu}$. Chosen scanning  range of  $t_{\beta}$ also satisfies the perturbative limit mentioned above.   

We comment here the results  obtained  previously in Ref. \cite{Hue:2021xap}, where large $t_{\beta}\ge 50$ and small values of  singly charged Higgs bosons $h^\pm_k$ ($k=1,2$) are required for  large $(g-2)_{\mu}$ satisfying $1\sigma$ experimental data of $(g-2)_{\mu}$ and all constraints from cLFV  decays $e_b\to e_a\gamma$. But only the case of $s_{\alpha}=0$ and non-zero $Y^{\sigma}_{22,33,23,32}$  was mentioned. Our numerical investigation shows that this case results in small $\Delta a_{e}$ which cannot satisfy $1\sigma$ range of  the experimental data given in Eq. \eqref{eq_dae}. Without  $\sigma^{\pm}$,  we  obtain  two maximal values of $\Delta a_{e}$ that $\Delta a_{e}\leq 2.5 \times 10^{-14}$ and $1.5 \times 10^{-14}$ for the NO and IO schemes, respectively. Hence,  determining the regions of parameter space giving large $\Delta a_e$ will be very interesting.

Because of the above reasons, we focus on the regions of parameter space giving large $\Delta a_{e}$  that satisfies   the $1\sigma$ experimental data of $(g-2)_e$ as well as all current constraints of cLFV decay rates  Br$(e_b\to e_a\gamma)$. The investigation  shows that the $1\sigma$  range of $\Delta a_e\in [1.8\times10^{-13}, 7.8\times 10^{-13}]$ can be obtained easily in a wide range of the parameter space, for example  the following  fixed values of  $ z_0=500$ GeV, $t_{\beta} =50$, and $s_{\alpha}=0.5$, and scanning the remaining parameters,  we have a benchmark point that  
 $m_{h^\pm_1}= 814.8$ GeV, $m_{h^\pm_2}= 771.5$ GeV, $m_{n_4}=m_{n_7}=2.152 $ TeV, $m_{n_5}=m_{n_8}= 4.365$ TeV,   $m_{n_6}=m_{n_9}=3.156$ TeV,  $s^{r}_{12}= -0.075$, $s^{r}_{13}=-0.565$, $s^{r}_{23}= -0.063$,  and  $0\leq |Y^\sigma_{ab}|\leq 0.293$,  which  results in the  following allowed  values of the relevant physical processes:
 $\Delta a_{e}= 4.243 \times 10^{-13}$,  $\Delta a_{\mu}= 1.019\times 10^{-9}$,  Br$(\mu\to e\gamma)= 2.95 \times 10^{-13}$, Br$(\tau\to e\gamma)=6.18 \times 10^{-9}$, Br$(\tau\to  \mu \gamma)= 3.52 \times 10^{-8}$, Br$(h^0_1\to \mu e)= 1.59\times 10^{-7}$,   Br$(h^0_1 \to \tau e)= 6.56 \times 10^{-6}$, and Br$(h^0_1\to \tau \mu )= 2.7 \times 10^{-4}$.

We list here other interesting benchmark points of the parameter space  corresponding to large $t_{\beta}=60$, that satisfy $1\sigma$ range of $(g-2)_e$, $\Delta a_{\mu}\geq 0.6\times 10^{-9}$, and all current LFV upper bounds. For other large $t_{\beta}$ values, the results are the same. 
\begin{enumerate}
		\item A benchmark point giving large Br$(h^0_1\to\tau e)\sim \mathcal{O}(10^{-5})$:
	\begin{align*}
	\{ z_0[\mathrm{GeV}],\; t_{\beta},\; s_{\alpha}\}&=\{867.7,\; 60,\; 0.460\},\; \{s^r_{12,13,23} \}= \{ 0.377,\; 0.556,\;-0.907  \},\crn
	\{m_{h^\pm_{1,2}}[\mathrm{TeV}]\}=& \{ 0.974,\; 0.918\}, \; \{m_{4,5,6}=m_{7,8,9} [\mathrm{TeV}]\}=\{ 3.32 ,\; 5.265,\; 3.341 \}, \crn
	Y^{\sigma}&= \begin{pmatrix}
	0.015& 0.006  & -0.013 \\ 
	-0.044 & 0.047 & -0.108 \\ 
	 0.003 & -0.183  & 0.063   
	\end{pmatrix}. 
	\end{align*}
	The corresponding  values of  $\Delta a_{e,\mu}$ and LFV decay rates are 
	\begin{align*}
	& \Delta a_e= 5.89 \times 10^{-13},\; \Delta a_\mu= 1.077 \times 10^{-9},   \crn 
	&  \mathrm{ Br}\{(\mu\to e\gamma),\; (\tau\to e\gamma),\; (\tau\to \mu\gamma)\}=\{8.31 \times 10^{-14}, \; 1.28 \times 10^{-8},\; 4.04 \times 10^{-8}\}\crn 
	&\; \mathrm{ Br}(h^0_1\to \{\mu e,\; \tau e,\; \tau \mu\})=\{ 5.9 \times 10^{-7},\; 5.91 \times 10^{-5},\; 5.18  \times 10^{-4}\}.
	\end{align*}
	
\item   There exists benchmark point that allows  large Br$(h^0_1\to\tau e)\sim \mathcal{O}(10^{-5})$, but small Br$(h^0_1\to\tau \mu)< \mathcal{O}(10^{-7})$: 
\begin{align*}
	&\{ z_0[\mathrm{GeV}],\; t_{\beta},\; s_{\alpha}\}=\{ 478.5 ,\; 60 ,\; 0.993 \},\; \{s^r_{12,13,23} \}= \{  0.629,\;  -0.867,\; -0.818 \},\crn
	&\{m_{h^\pm_{1,2}}[\mathrm{TeV}]\}= \{0.997,\; 0.864 \}, \; \{m_{4,5,6}=m_{7,8,9} [\mathrm{TeV}]\}=\{ 2.994,\; 4.092,\;2.378  \}, \crn
	&Y^{\sigma} = \begin{pmatrix}
		0.069 & 0.164  & -0.085  \\ 
		 -0.076 & 0.058  & -0.199  \\ 
		0.074 & -0.180    & -0.086  
	\end{pmatrix},\crn 
	& \Delta a_e= 4.67 \times 10^{-13},\; \Delta a_\mu= 0.998  \times 10^{-9},   \crn 
	& \mathrm{ Br}\{(\mu\to e\gamma),\; (\tau\to e\gamma),\; (\tau\to \mu\gamma)\}=\{ 2.196 \times 10^{-13}, \;  3.523 \times 10^{-9},\;  3.547\times 10^{-8}\}, \crn 
	&\;\mathrm{ Br}(h^0_1\to \{\mu e,\; \tau e,\; \tau \mu\})=\{  5.93\times 10^{-6},\; 1.88 \times 10^{-5},\;  6.56 \times 10^{-8}\}.
\end{align*}

	\item There exists  a benchmark point  predicting  large Br$(h\to e\mu)\sim \mathcal{O}(10^{-5})$, which is close to the experimental constraint: 
\begin{align*}
	&\{ z_0[\mathrm{GeV}],\; t_{\beta},\; s_{\alpha}\}=\{1019.5 ,\; 60,\; 0.848  \},\; \{s^r_{12,13,23} \}= \{ 0.11,\; -0.89,\; -0.822   \},\crn
	&\{m_{h^\pm_{1,2}}[\mathrm{TeV}]\}= \{ 0.671,\;0.622   \}, \; \{m_{4,5,6}=m_{7,8,9} [\mathrm{TeV}]\}=\{ 6.533,\;9.657 ,\;4.414   \}, \crn
	&Y^{\sigma} = \begin{pmatrix}
		0.079 & 0.189  & -0.113   \\ 
		-0.094& -0.061   & -0.210  \\ 
	0.079	 & -0.241    &-0.059   
	\end{pmatrix},\crn 
	& \Delta a_e= 3.19 \times 10^{-13},\; \Delta a_\mu= 0.917 \times 10^{-9},   \crn 
	& \mathrm{ Br}\{(\mu\to e\gamma),\; (\tau\to e\gamma),\; (\tau\to \mu\gamma)\} =\{ 2.37 \times 10^{-13}, \;  2.80  \times 10^{-9},\; 3.07   \times 10^{-8}\}, \crn 
	&\;\mathrm{ Br}(h^0_1\to \{\mu e,\; \tau e,\; \tau \mu\})=\{ 1.85\times 10^{-5},\; 1.05 \times 10^{-4},\;  1.48  \times 10^{-3}\}.
\end{align*}
It is noted that large Br$(h\to e\mu)$ requires both large $z_0$ and Br$(h\to \tau\mu)\sim \mathcal{O}(10^{-3})$ which may be excluded by planned experiments.  In this case, the numerical results show that  Br$(h\to \tau\mu)< \mathcal{O}(10^{-4})$ will lead to  Br$(h\to e\mu)< \mathcal{O}(10^{-6})$, which is still smaller than the planned experimental sensitivity.  
\end{enumerate}

From our numerical investigation, we  found that  the regions allowing $1\sigma$ range of $\Delta a_e$ data and cLFV constraints are very wide. But the regions allowed large $(g-2)_{\mu}$ are difficult to control. This is because of the large number of free parameters in the 331ISS  that our numerical code is not still smart enough to collect these points.   Because of the special form of $\tilde{m}_D$ that require the non-degenerate matrix $\hat{k}$ and the strong destructive correlations between the mixing angles  $s^r_{ab}$ and the entries of  $Y^\sigma$ in order to get small Br$(e_b\to e_a\gamma)$, in the regions allow large $\Delta a_{\mu}$. There may exist some certain relations between these parameters for collecting more interesting points allowing large $(g-2)_{\mu}$ at $1\sigma$ experimental range.   We will determine them in a future work. 

 Finally,  we comment some properties of the current $Z$ boson decay data which may put useful constraints on the parameter space of the 331ISS model.  In the limit of $v/w\to0$, equivalently $t_{\theta}=0$,  the couplings of $Z$ boson with all other SM particles.  We can see that all masses of the  new heavy neutrinos  appearing in the  collected points we showed above as  the numerical results are much larger than the $Z$ boson masses. Therefore, $Z$ do have not any new tree level decays $Z\to\overline{n}_I n_j $ with at least a new heavy neutrino $n_I$ ($I>3$).  In addition, all masses of the new heavy particles predicted by the 331ISS models are heavier than the $Z$ boson masses, therefore the invisible decays of the $Z$ boson in this case is the same as that in the SM and the 2HDM  discussed in Ref. \cite{Jurciukonis:2021izn}. We therefore conclude that the  current $Z$ boson decay data affects weakly the allowed region of the parameters space we focus on this work.  

There is another cLFV decay mode $Z\to e_a^+e^-_b$ discussed in detailed in 2HDM \cite{Jurciukonis:2021izn}, which is still invisible in the regions predicting large Br$(h\to e_ae_b)$ and satisfying all the constraints of cLFV decays Br$(e_b\to e_a\gamma)$. Therefore, this decay channel will not change significantly the allowed regions of parameters discussed in this work. On the other hand, the interesting topic we will  focus on is that when the experimental sensitivities are improved, both cLFV decays of $\mu^-\to e^-\overline{\nu}_e\nu_{\mu} $ and $Z\to e_ae_b$ may give more significant constraints on those mentioned in this work.

\section{\label{conclusion} Conclusion}
In this work, we have constructed  the  analytic formulas  for  one-loop contributions to the LFV decays of the SM-like Higgs boson $h^0_1\to e_ae_b$ in the 331ISS model. We also  give analytic formulas to explain qualitatively  the results of large $(g-2)_{\mu}$ reported previously. Numerical tests were  used to confirm the consistency between the two calculations.  We introduced a new parameterization of the heavy neutrino mass matrix to reduce the number of free parameters used to investigate $(g-2)_{e,\mu}$ anomalies,  LFV decays $e_b\to e_a\gamma $,  and $h^0_1\to e_a e_b$.  Our numerical investigation shows that the model can predict easily the $1\sigma$ range  of  experimental data for $(g-2)_e$ and satisfy simultaneously the cLFV constrains Br$(e_b\to e_a\gamma)$. But we only obtained  the regions of parameter space that give largest values of  $\Delta a_{\mu}\simeq 10^{-9}$, which rather smaller than the lower bound of $1\sigma$ range reported recently.  The  reason is that the  recent numerical code used in our investigation only works  in the limit of small max$[|y^{\sigma}|]<0.25$. In these regions of the parameter space, the largest values of Br$(h^0_1\to \tau e)$ and Br$(h^0_1\to \tau \mu)$ are order of $\mathcal{O}(10^{-4})$ and $10^{-3}$, respectively. In addition,  large Br$(h^0_1\to \tau \mu)$ predicts large  Br$(h^0_1\to \mu e) \sim\mathcal{O}(10^{-5})$, which is close to the recent experimental bounds.   The regions  with  large $|Y^\sigma|_{ab}$ may be more interesting, which is our future work, where many other LFV processes such as $Z\to e_be_a$, $e_b\to e_{c}e_{d}e_{f}$, and the $\mu-e$ conversion in nuclei will be discussed together.

\section*{Acknowledgments}
We are grateful Prof. Martin Hofericher, and  Dr. Mukesh Kumar   for their communications. We would like to  thank the referee for reminding us of the important contribution of the singly charged Higgs bosons to the decay  $\mu^-\to e^- \overline{\nu_e}\nu_{\mu}$, which significantly changes   our numerical results. This research is funded by An Giang University under grant number 21.01.TB. L. T. Hue is grateful  to Van Lang University.

\appendix
\section{\label{DeltaLR}Form factors of LFVH in the unitary gauge}
The one-loop contributions here are calculated using the  notations of  Passarino-Veltman (PV)  functions \cite{tHooft:1972tcz, Denner:2005nn} given in Ref.~\cite{Nguyen:2020ehj}, consistent with LoopTools \cite{Hahn:1998yk},  see a detailed discussion in Refs. \cite{Hue:2015fbb, Phan:2016ouz}.  The  PV functions used in this work are defined as follows: $B^{(i)}_{\mu}\equiv B_1^{(i)}\times (-1)^i p_{i\mu}$ with $i=1,2$, and $C_{\mu}\equiv \sum _{i=1}^2  (-1)^i p_{i\mu}\times C_i$. As mentioned in Ref. \cite{Nguyen:2020ehj}, the two $B_1^{(1)}$ and $C_1$ have opposite signs with those introduced in Ref. \cite{Hue:2015fbb}. They come from the signs of $p_{1,2}$ in the internal momenta $(k-p_1)$ and   $(k+p_2)$ shown in Fig. \ref{hlilj1}, which $p_1$ has an opposite sign, which is  different from   the standard notation of $k+p_{1}$ defined in LoopTools. The PV-functions used  in our formulas are: $B^{(i)}_{0,1}=B_{0,1}(p_i^2; M^2_0,M^2_i)$, $C_{0,1,2}=C_{0,1,2}(p_1^2,(p_1+p_2)^2, p_2^2; M_0^2,M_1^2,M_2^2)$, and  $B^{(12)}_0=B_0((p_1+p_2)^2; M^2_1,M^2_2)$. In the below, when the external momenta are fixed as  $p_1^2=m^2_{e_a}$, $p_2^2=m^2_{e_b}$, and $(p_1+p_2)^2=m_{h^0_1}^2$, we use the simpler  notations  as follows $C_{0,1,2}(p_1^2, m^2_{h^0_1}, p_2^2; M_0^2,M_1^2,M_2^2)\equiv C_{0,1,2}( M_0^2,M_1^2,M_2^2)$, $B^{(i)}_{0,1}(M^2_0,M^2_i)=B_{0,1}(p_i^2; M^2_0,M^2_i)$, and $B_0(m_{h^0_1}^2; M^2_1,M^2_2)=B^{(12)}_0( M^2_1,M^2_2)$. 

The analytic expressions $\Delta^{(i)W}_{L,R}\equiv \Delta^{(i)W}_{(ab)L,R}$ for   one-loop contributions from the diagram (i) in Fig. \ref{hlilj1}  are 
\begin{align}
	\Delta^{(1)W}_{L} &= \frac{g^3 m_{a}}{64\pi^2 m_W^3} \sum_{i=1}^{9}U^{\nu}_{ai}U^{*\nu}_{bi}
	\left\{ m_{n_i}^2\left( B^{(1)}_0 + B^{(2)}_0 +B^{(1)}_1\right) +m_b^2 B^{(2)}_1  - \left(2m_W^2+m^2_{h^0_1}\right)m_{n_i}^2 C_0 \right.\crn &-\left. \left[m_{n_i}^2 \left( 2m_W^2 +m_{h^0_1}^2  \right)+ 2m_W^2\left(2m_W^2 +m_a^2 -m_b^2\right)  \right] C_1 \right. 
	\crn -&\left.  
	\left[2m_W^2\left(m_a^2-m^2_{h^0_1}\right)+ m_b^2 m^2_{h^0_1}\right]C_2\frac{}{}\right\},\crn
	\Delta^{(1)W}_{R}&=  \frac{g^3 m_{b}}{64\pi^2 m_W^3} \sum_{i=1}^{9}U^{\nu}_{ai}U^{*\nu}_{bi}
	\left\{ m_{n_i}^2\left( B^{(1)}_0 + B^{(2)}_0 +B^{(2)}_1\right) +m_a^2 B^{(1)}_1  - \left(2m_W^2+m^2_{h^0_1}\right)m_{n_i}^2 C_0 \right.\crn &-\left. \left[m_{n_i}^2 \left( 2m_W^2 +m_{h^0_1}^2  \right)+ 2m_W^2\left(2m_W^2 -m_a^2 +m_b^2\right)  \right] C_2 \right. 
	\crn -&\left.  
	\left[2m_W^2\left(m_b^2-m^2_{h^0_1}\right)+ m_a^2 m^2_{h^0_1}\right]C_1\frac{}{}\right\},\crn
\Delta^{(7+8)W}_{L} &=  \frac{g^3m_am_b^2}{64\pi^2m^3_W(m_b^2-m_a^2)} 
\crn &\times \sum_{i=1}^{9}U^{\nu}_{ai}U^{\nu*}_{bi}\left[  2m_{n_i}^2\left(B^{(2)}_0-B^{(1)}_0\right) +  \left(2 m_W^2 +m_{n_i}^2\right) \left(B^{(2)}_1-B^{(1)}_1 \right) +m_b^2 B^{(2)}_1 - m_a^2 B^{(1)}_1  \right],  
\crn
\Delta^{(7+8)W}_{R} &=\frac{m_a}{m_b}\Delta^{(7+8)W}_{L}, \nn 
\end{align}
where $B^{(k)}_{0,1}=B^{(k)}_{0,1}(m^2_{n_i},m_W^2)$ and $C_{0,1,2}=C_{0,1,2}(m^2_{n_i},m_W^2,m_W^2)$,
\begin{align}
	\Delta^{(5)W}_{L}=& \frac{g^3m_a}{64\pi^2 m_W^3}\sum_{i,j=1}^9U^{\nu*}_{ai}U^{\nu}_{bj}
	\left\{D_{ij}\left[-m_{n_j}^2 B^{(12)}_0 + m_{n_i}^2 B^{(1)}_1 +  m^2_{n_j}m_{W}^2 C_0\right.\right. \crn
	+& \left.  \left(2 m_W^2 (m_{n_i}^2+ m_{n_j}^2)  +2 m_{n_i}^2m_{n_j}^2 -m_a^2 m_{n_j}^2 -m_b^2 m_{n_i}^2\right) C_1 \right]\crn
	+&\left. D^*_{ij}m_{n_i}m_{n_j}\left[ -B^{(12)}_0 + B^{(1)}_1 +m_{W}^2 C_0+ \left(4 m_W^2 + m_{n_i}^2+m_{n_j}^2-m_a^2  -m_b^2\right) C_1\right]   \right\}, \crn
	%
	\Delta^{(5)W}_{R}=& \frac{g^3m_b}{64\pi^2 m_W^3}\sum_{i,j=1}^9U^{\nu*}_{ai}U^{\nu}_{bj} \left\{D_{ij}\left[-m_{n_i}^2 B^{(12)}_0 + m_{n_j}^2 B^{(2)}_1 +  m^2_{n_i}m_{W}^2 C_0\right.\right. \crn
	+&\left.  \left(2 m_W^2 (m_{n_i}^2+ m_{n_j}^2) +2 m_{n_i}^2m_{n_j}^2 -m_a^2 m_{n_j}^2  -m_b^2 m_{n_i}^2\right) C_2 \right]\crn
	+ & \left. D^*_{ij}m_{n_i}m_{n_j}\left[ -B^{(12)}_0  + B^{(2)}_1 + m_{W}^2 C_0 +\left(4 m_W^2 + m_{n_i}^2+m_{n_j}^2-m_a^2 -m_b^2 \right) C_2 \right]\right\},\nn 
\end{align}
where $D_{ij}=\sum_{c=1}^3 U^{\nu}_{ci}U^{\nu*}_{cj}$,   $B^{(12)}_{0}=B^{(12)}_{0}(m^2_{n_i},m^2_{n_j})$,   $B^{(1)}_{1}=B^{(1)}_{1}(m^2_{W},m^2_{n_i})$, $B^{(2)}_{1}=B_{1}^{(2)}(m^2_{W},m^2_{n_j})$, and $C_{0,1,2}=C_{0,1,2}(m^2_{W},m_{n_i}^2,m_{n_j}^2)$. 
The analytic expressions $\Delta^{(i)Y}_{L,R}\equiv  \Delta^{(i)Yh^{\pm}_3}_{(ab)L,R}$ with $i=4,6,9,10$,  are
\begin{align}
	 \Delta^{(1)Y}_{L} &= \frac{g^3  m_{a}c_{\beta}s_{\theta}}{64 \pi^2  m_Y^3} \sum_{i=1}^{9}U^{\nu}_{(a+3)i}U^{\nu*}_{(b+3)i}
	\left\{ m_{n_i}^2\left( B^{(1)}_0+  B^{(2)}_0 +B^{(1)}_1\right) +m_b^2 B^{(2)}_1\right.\crn
	& -\left.\left(2m_Y^2+m^2_{h^0_1}\right)m_{n_i}^2 C_0 -  \left[2m_Y^2\left(2m_Y^2 +m_a^2-m_b^2\right)  + m_{n_i}^2 \left( 2m_Y^2 +m_{h^0_1}^2 \right)\right] C_1 \right.\crn
	&+\left.
	\left[2m_Y^2\left(m_a^2-m^2_{h^0_1}\right)+ m_b^2 m^2_{h^0_1}\right]C_2\frac{}{}\right\},\crn
	\Delta^{(1)Y}_{R} &= \frac{g^3  m_{b}c_{\beta}s_{\theta}}{64 \pi^2  m_Y^3} \sum_{i=1}^{9}U^{\nu*}_{(a+3)i}U^{\nu}_{(b+3)i}
	\left\{ m_{n_i}^2\left( B^{(1)}_0+  B^{(2)}_0 +B^{(2)}_1\right) +m_a^2 B^{(1)}_1\right.\crn
	& -\left.\left(2m_Y^2+m^2_{h^0_1}\right)m_{n_i}^2 C_0 -  \left[2m_Y^2\left(2m_Y^2 -m_a^2 +m_b^2\right)  + m_{n_i}^2 \left( 2m_Y^2 +m_{h^0_1}^2 \right)\right] C_2 \right.\crn
	&+\left.
	\left[2m_Y^2\left(m_b^2-m^2_{h^0_1}\right)+ m_b^2 m^2_{h^0_1}\right]C_1\frac{}{}\right\},\crn
	\Delta^{(7+8)Y}_{L} &= \frac{g^3m_am_b^2}{64\pi^2m_Wm_Y^2(m_b^2 -m_a^2)}  \sum_{i=1}^{9}U^{\nu*}_{(a+3)i}U^{\nu}_{(b+3)i} \crn &\times \left[  2m_{n_i}^2\left(B^{(2)}_0-B^{(1)}_0\right) +\left(2 m_Y^2 +m_{n_i}^2\right) \left(B^{(2)}_1 -B^{(1)}_1 \right) - m_a^2 B^{(1)}_1 +m_b^2 B^{(2)}_1 \right],  \label{d78YL}\crn
	\Delta^{(7+8)Y}_{R} &=\frac{m_a}{m_b}\Delta^{(7+8)Y}_{L},\nn
\end{align}
where  $B^{(k)}_{0,1}=B^{(k)}_{0,1}(m^2_{n_i},m_Y^2)$ and $C_{0,1,2}=C_{0,1,2}(m^2_{n_i},m_Y^2,m_Y^2)$. One-loop contributions from diagram 5 are 
\begin{align}
\Delta^{(5)Y}_{L} &=\frac{g^3 m_a}{64\pi^2  m_Wm_Y^2}\crn
&\times \sum_{i,j=1}^{9}U^{\nu}_{(a+3)i}U^{\nu*}_{(b+3)j} \left\{D_{ij}\left[-m_{n_j}^2 B^{(12)}_0 + m_{n_i}^2 B^{(1)}_1 +  m^2_{n_j}m_{W}^2 C_0\right.\right. \crn
+& \left.  \left(2 m_W^2 (m_{n_i}^2+ m_{n_j}^2)  +2 m_{n_i}^2m_{n_j}^2 -m_a^2 m_{n_j}^2 -m_b^2 m_{n_i}^2\right) C_1 \right]\crn
+&\left. D^*_{ij}m_{n_i}m_{n_j}\left[ -B^{(12)}_0 + B^{(1)}_1 +m_{W}^2 C_0+ \left(4 m_W^2 + m_{n_i}^2+m_{n_j}^2-m_a^2  -m_b^2\right) C_1\right]   \right\} ,\crn
\Delta^{(5)Y}_{R} &= \frac{g^3 m_b}{64\pi^2  m_Wm_Y^2} \crn
&\times \sum_{i,j=1}^{9}U^{\nu}_{(a+3)i}U^{\nu*}_{(b+3)j} \left\{D_{ij}\left[-m_{n_i}^2 B^{(12)}_0 + m_{n_j}^2 B^{(2)}_1 +  m^2_{n_i}m_{W}^2 C_0\right.\right. \crn
+&\left.  \left(2 m_W^2 (m_{n_i}^2+ m_{n_j}^2) +2 m_{n_i}^2m_{n_j}^2 -m_a^2 m_{n_j}^2  -m_b^2 m_{n_i}^2\right) C_2 \right]\crn
+ & \left. D^*_{ij}m_{n_i}m_{n_j}\left[ -B^{(12)}_0  + B^{(2)}_1 + m_{W}^2 C_0 +\left(4 m_W^2 + m_{n_i}^2+m_{n_j}^2-m_a^2 -m_b^2 \right) C_2 \right]\right\},\nn 
\end{align}
where $B^{(12)}_{0}=B^{(12)}_{0}(m^2_{n_i},m^2_{n_j})$,   $B^{(1)}_{1}=B^{(1)}_{1}(m^2_{Y},m^2_{n_i})$, $B^{(2)}_{1}=B_{1}^{(2)}(m^2_{Y},m^2_{n_j})$, and $C_{0,1,2}=C_{0,1,2}(m^2_{Y},m_{n_i}^2,m_{n_j}^2)$. 
\begin{align}
\Delta^{(2)Y}_{L} &= -\frac{g^3  m_{a}c_{\theta}c_\beta}{64 \pi^2 m_W m_Y^2}\sum_{i=1}^{9}U^{\nu}_{(a+3)i}\crn
&\times \left\{\lambda^{L,1}_{bi}m_{n_i}\left[ B^{(1)}_0 +B^{(1)}_1+\left(m_Y^2+m_{h^\pm_3}^2-m_{h^0_1}^2\right)C_0 - \left(m_Y^2-m_{h^\pm_3}^2+m_{h^0_1}^2\right)C_1\right]\right.\crn
& -\left. \lambda^{R,1}_{bi}m_{b}\left[ 2m_Y^2C_1 +\left(m_Y^2+m_{h^\pm_3}^2-m_{h^0_1}^2\right)C_2\right]\right\},\crn
\Delta^{(2)Y}_{R} &= \frac{g^3  c_{\theta}c_\beta}{64 \pi^2 m_W m_Y^2}\sum_{i=1}^{9}U^{\nu}_{(a+3)i}\crn
&\times \left\{\lambda^{L,1}_{bi}m_bm_{n_i}\left[ 2m_Y^2C_0 +\left(m_Y^2-m_{h^\pm_3}^2+m_{h^0_1}^2\right)C_2\right]\right.\crn
&+
\left. \lambda^{R,1}_{bi}\left[ m_{n_i}^2 B^{(1)}_0 +m_a^2B^{(1)}_1 -m_{n_i}^2\left(m_Y^2-m_{h^\pm_3}^2+m_{h^0_1}^2\right)C_0\right.\right.\crn
& +\left.\left.\left[ 2m_Y^2\left(m_{h^0_1}^2-m_b^2\right)- m_a^2\left(m_Y^2-m_{h^\pm_3}^2+m_{h^0_1}^2\right)\right]C_1 -2 m_b^2m_Y^2C_2\right]\right\},\nn
\end{align}
where  $B^{(1)}_{k}=B^{(1)}_{k}(m^2_{Y},m^2_{n_i})$ ($k=0,1$) and $C_{0,1,2}=C_{0,1,2}(m_{n_i}^2, m^2_{Y}m_{h^\pm_3}^2)$, 
\begin{align}
\Delta^{(3)Y}_{L} &= \frac{g^3  c_{\theta}c_\beta}{64 \pi^2 m_W m_Y^2}\sum_{i=1}^{9}U^{\nu*}_{(b+3)i}\crn
&\times \left\{\lambda^{L,1*}_{ai}m_am_{n_i}\left[  2m_Y^2C_0+\left(m_Y^2-m_{h^\pm_3}^2+m_{h^0_1}^2\right)C_1\right]\right.\crn
&+
\left. \lambda^{R,1*}_{ai}\left[ m_{n_i}^2 B^{(2)}_0 +m_b^2B^{(2)}_1 - m_{n_i}^2\left(m_Y^2-m_{h^\pm_3}^2+m_{h^0_1}^2\right)C_0\right.\right.\crn
& -\left.\left. 2m_a^2m_Y^2C_1 +\left[ 2m_Y^2\left(m_{h^0_1}^2-m_a^2\right) - m_b^2\left(m_Y^2-m_{h^\pm_3}^2+m_{h^0_1}^2\right)\right]C_2\right]\right\},\crn
\Delta^{(3)Y}_{R} &= -\frac{g^3  m_{b}c_{\theta}c_\beta}{64 \pi^2 m_W m_Y^2} 
\crn&\times  \sum_{i=1}^{9}U^{\nu*}_{(b+3)i}  \left\{\lambda^{L,1*}_{ai}m_{n_i}\left[ B^{(2)}_0+B^{(2)}_1
+\left(m_Y^2+m_{h^\pm_3}^2-m_{h^0_1}^2\right)C_0-\left(m_Y^2-m_{h^\pm_3}^2+m_{h^0_1}^2\right)C_2\right]\right.\crn
& - \left. \lambda^{R,1*}_{ai}m_{a}\left[  \left(m_Y^2+m_{h^\pm_3}^2-m_{h^0_1}^2\right)C_1 +2m_Y^2C_2\right]\right\},\nn 
\end{align}
where  $B^{(2)}_{k}=B^{(2)}_{k}(m^2_{Y},m^2_{n_i})$ and $C_{0,1,2}=C_{0,1,2}(m_{n_i}^2,m_{h^\pm_3}^2, m^2_{Y})$, 
\begin{align}
	\Delta^{(4)h^{\pm}_{k,l}}_{L}&= \frac{g^2g_{hkl}}{32\pi^2  m_W^2}\sum_{i=1}^{9}\left[-\lambda^{R,k*}_{ai}\lambda^{L,k}_{bi}m_{n_i}C_0 +\lambda^{L,k*}_{ai}\lambda^{L,k}_{bi}m_{a}C_1 +\lambda^{R,k*}_{ai}\lambda^{R,k}_{bi}m_{b}C_2 \right],\crn
	\Delta^{(4)h^{\pm}_{k,l}}_{R} &=\frac{g^2g_{hkl}}{32\pi^2 m_W^2} \sum_{i=1}^{9}\left[-\lambda^{L,k*}_{ai}\lambda^{R,k}_{bi}m_{n_i}C_0 +\lambda^{R,k*}_{ai}\lambda^{R,k}_{bi}m_{a} C_1 +\lambda^{L,k*}_{ai}\lambda^{L,k}_{bi}m_{b}C_2 \right],	\nn 
\end{align}
where $\{k,l\}=\{1,2\},\{2,1\}, \{1,1\},\{2,2\},\{3,3\}$, $g_{h21}=g_{h12}$  and $C_{0,1,2}=C_{0,1,2}(m_{n_i}^2,m_{h^\pm_k}^2,m_{h^\pm_l}^2)$, 
 \begin{align}
 \Delta^{(6)h^{\pm}_k}_{L} &=    \frac{g^3}{64\pi^2  m_W^3} \sum_{i,j=1}^{9}\left\{\lambda^{0*}_{ij}\left[\lambda^{R,k*}_{ai}\lambda^{L,k}_{bj}\left(B^{(12)}_0+m_{h^\pm_k}^2C_0 +m_a^2 C_1 +m_b^2C_2\right)\right.\right.\crn
 &+\left.\left. \lambda^{R,k*}_{ai}\lambda^{R,k}_{bj}m_bm_{n_j}C_2 +\lambda^{L,k*}_{ai}\lambda^{L,k}_{bj}m_am_{n_i}C_1 \right]\right. \crn
 &+\left.  \lambda^{0}_{ij}\left[\lambda^{R,k*}_{ai}\lambda^{L,k}_{bj}m_{n_i}m_{n_j}C_0 +\lambda^{R,k*}_{ai}\lambda^{R,k}_{bj}m_{n_i}m_{b}(C_0+C_2)\right.\right.\crn
 &+\left.\left.\lambda^{L,k*}_{ai}\lambda^{L,k}_{bj}m_{a}m_{n_j}(C_0 +C_1)+ \lambda^{L,k*}_{ai}\lambda^{R,k}_{bj}m_{a}m_{b}(C_0 +C_1 +C_2) \right]\frac{}{}\right\},\crn
\Delta^{(6) h^{\pm}_k}_{R} &= \frac{g^3}{64\pi^2  m_W^3} \sum_{i,j=1}^{9}\left\{\lambda^{0}_{ij}\left[\lambda^{L,k*}_{ai}\lambda^{R,k}_{bj}\left(B^{(12)}_0+m_{h^\pm_k}^2C_0  +m_a^2 C_1 +m_b^2C_2\right)\right.\right.\crn
&+\left.\left.\lambda^{L,k*}_{ai}\lambda^{L,k}_{bj}m_bm_{n_j}C_2 +  \lambda^{R,k*}_{ai}\lambda^{R,k}_{bj}m_am_{n_i}C_1 \right]\right.\crn
&+\left.
\lambda^{0*}_{ij}\left[\lambda^{L,k*}_{ai}\lambda^{R,k}_{bj}m_{n_i}m_{n_j}C_0 +\lambda^{L,k*}_{ai}\lambda^{L,k}_{bj}m_{n_i}m_{b}(C_0+C_2)\right.\right.\crn
&+ \left.\left.\lambda^{R,k*}_{ai}\lambda^{R,k}_{bj}m_{a}m_{n_j}(C_0 +C_1)+ \lambda^{R,k*}_{ai}\lambda^{L,k}_{bj}m_{a}m_{b}(C_0+ C_1+C_2) \right]\frac{}{}\right\}, \nn 
 \end{align}
where $k=1,2,3$,  $B^{(12)}_0=B^{(12)}_0(m^2_{n_i}, m^2_{n_j})$, and $C_{0,1,2}=C_{0,1,2}(m_{n_i}^2,m_{n_j}^2,m_{h^\pm_k}^2)$, 
\begin{align}
\Delta^{(9+10)h^{\pm}_k}_{L}  &= \frac{g^3}{64\pi^2m_W^3\left(m_a^2-m_b^2\right)} \crn
&\times\sum_{i=1}^{9}\left[ m_am_bm_{n_i}  \lambda^{L,k*}_{ai}\lambda^{R,k}_{bi}\left(B^{(1)}_0-B^{(2)}_0\right)+ m_{n_i}
\lambda^{R,k*}_{ai}\lambda^{L,k}_{bi}\left(m^2_bB^{(1)}_0-m^2_aB^{(2)}_0\right)\right.\crn
&\left.+ m_{a}m_b \left(\lambda^{L,k*}_{ai}\lambda^{L,k}_{bi}m_b + \lambda^{R,k*}_{ai}\lambda^{R,k}_{bi}m_a\right)\left(-B^{(1)}_1+ B^{(2)}_1\right)\right],\crn
\Delta^{(9+10) h^{\pm}_k}_{R}  &=\frac{g^3}{64 \pi^2m_W^3\left(m_a^2-m_b^2\right)}  \crn
&\times\sum_{i=1}^{9}\left[ m_am_bm_{n_i}  \lambda^{R,k*}_{ai}\lambda^{L,k}_{bi}\left(B^{(1)}_0-B^{(2)}_0\right)+ m_{n_i}
\lambda^{L,k*}_{ai}\lambda^{R,k}_{bi}\left(m^2_bB^{(1)}_0-m^2_aB^{(2)}_0\right)\right.\crn
&\left.+ m_{a}m_b \left(\lambda^{R,k*}_{ai}\lambda^{R,k}_{bi}m_b + \lambda^{L,k*}_{ai}\lambda^{L,k}_{bi}m_a\right)\left( -B^{(1)}_1+ B^{(2)}_1\right)\right], \nn 
\end{align}
where $k=1,2,3$,  $B^{(k)}_{0,1}=B^{(k)}_{0,1}(m^2_{n_i},m^2_{h^\pm_k})$. 
The details to  derive the above formulas  of  $\Delta^{(i)}_{L,R}$ were shown in Refs. \cite{Thao:2017qtn, Boucenna:2015zwa}, and hence we do not present them in this work.  We note that  the scalar functions $\Delta^{(1)W}_{L,R}$  and $\Delta^{(1,2,3)Y}_{L,R}$ include  parts that do not depend on $m_{n_i}$, and therefore they vanish because of the Glashow-Iliopoulos-Maiani mechanism. 

The divergent cancellation in the total $\Delta_{L,R}$ is shown as follows. 
	\bea \mathrm{div}\left[\Delta^{(1)W}_L\right]&=&m_a\Delta_{\epsilon}\times  \frac{3}{2} \times \sum_{i=1}^9 U^{\nu*}_{ai}U^{\nu}_{bi} m^2_{n_i},\crn
\mathrm{div}\left[\Delta^{(5)W}_L\right]&=&m_a \Delta_{\epsilon}\times \sum_{i,j=1}^9 U^{\nu*}_{ai}U^{\nu}_{bj}\left(-D^{*}_{ij} m_{n_j}^2-\frac{1}{2}D_{ij} m_{n_i}^2\right),\crn
\mathrm{div}\left[\Delta^{(7+8)W}_L\right]&=&\mathrm{div}\left[\Delta^{(4)Y}_L\right]= \mathrm{div}\left[\Delta^{(7+8)Y}_L\right]=0,\crn
\mathrm{div}\left[\Delta^{(1)Y}_L\right]&=&m_a \Delta_{\epsilon}\times  \left(\frac{3s_\theta^4}{2c_\beta^2} \right) \sum_{i=1}^9 U^{\nu}_{(a+3)i}U^{\nu}_{(b+3)i} m^2_{n_i},\crn
\mathrm{div}\left[\Delta^{(2)Y}_L\right]&=&m_a\Delta_{\epsilon}\times \left( -\frac{c_\theta s_\theta^2}{2c_\beta} \right)\sum_{i=1}^9 U^{\nu}_{(a+3)i}\lambda^{L,1}_{bi} m_{n_i},\crn
\mathrm{div}\left[\Delta^{(3)Y}_L\right]&=&\Delta_{\epsilon}\times \left(\frac{c_\theta s_\theta^2}{c_\beta}\right)  \sum_{i=1}^9 U^{\nu*}_{(a+3)i}L_{ai}^{R,1} m^2_{n_i},\crn
\mathrm{div}\left[\Delta^{(5)Y}_L\right]&=&m_a \Delta_{\epsilon}\times \frac{s_\theta^2}{c_\beta^2} \sum_{i,j=1}^9 U^{\nu*}_{(a+3)i}U^{\nu}_{(b+3)j} \left(-D^{*}_{ij} m_{n_j}^2-\frac{1}{2}D_{ij} m_{n_i}^2\right),\crn
\mathrm{div}\left[\Delta^{(6)Yh_k^\pm}_L\right]&=&m_a\Delta_{\epsilon}\times  \sum_{i,j=1}^9 U^{\nu*}_{(a+3)i}\lambda^{0*}_{ij}\lambda^{L,k}_{bj},\crn
\mathrm{div}\left[\Delta^{(9+10)Yh_k^\pm}_L\right]&=&-m_a\Delta_{\epsilon}\times  \sum_{i=1}^9 U^{\nu*}_{(a+3)i}\lambda^{L,k}_{bi} m_{n_i},  
\label{div1}
\eea
where:  
div$B^{(1)}_0=$div$B^{(2)}_0=$div$B^{(12)}_0=-2$div$B^{(1)}_1=-2$ div$B^{(2)}_1=\Delta_{\epsilon}$ and $1/m_Y=s_{\theta}/{(c_\beta m_W)}$.

Easy to see that 
$\mathrm{div}\left[\Delta^{(1)W}_L\right]+\mathrm{div}\left[\Delta^{(5)W}_L\right]=\mathrm{div}\left[\Delta^{(6)Yh_k^\pm}_L\right]+\mathrm{div}\left[\Delta^{(9+10)Yh_k^\pm}_L\right]=0$  and the sum of the remaining divergent parts is zero in case we are focusing on investigating $c_\theta =1$.


\begin{thebibliography}{9}
\bibitem{Super-Kamiokande:1998kpq}
Y.~Fukuda \textit{et al.} [Super-Kamiokande],
Phys. Rev. Lett. \textbf{81} (1998), 1562-1567
[arXiv:hep-ex/9807003 [hep-ex]].

\bibitem{Super-Kamiokande:2001ljr}
S.~Fukuda \textit{et al.} [Super-Kamiokande],
Phys. Rev. Lett. \textbf{86} (2001), 5651-5655
[arXiv:hep-ex/0103032 [hep-ex]].

\bibitem{Super-Kamiokande:2001bfk}
S.~Fukuda \textit{et al.} [Super-Kamiokande],
Phys. Rev. Lett. \textbf{86} (2001), 5656-5660
[arXiv:hep-ex/0103033 [hep-ex]].

\bibitem{SNO:2002tuh}
Q.~R.~Ahmad \textit{et al.} [SNO],
Phys. Rev. Lett. \textbf{89} (2002), 011301
[arXiv:nucl-ex/0204008 [nucl-ex]].

\bibitem{SNO:2002hgz}
Q.~R.~Ahmad \textit{et al.} [SNO],
Phys. Rev. Lett. \textbf{89} (2002), 011302
[arXiv:nucl-ex/0204009 [nucl-ex]].


\bibitem{BaBar:2009hkt}
B.~Aubert \textit{et al.} [BaBar],
Phys. Rev. Lett. \textbf{104}, 021802 (2010)
[arXiv:0908.2381 [hep-ex]].

\bibitem{MEG:2016leq}
A.~M.~Baldini \textit{et al.} [MEG],
Eur. Phys. J. C \textbf{76}, no.8, 434 (2016)
[arXiv:1605.05081 [hep-ex]].

\bibitem{Belle-II:2018jsg}
E.~Kou \textit{et al.} [Belle-II],
PTEP \textbf{2019}, no.12, 123C01 (2019)
[erratum: PTEP \textbf{2020}, no.2, 029201 (2020)]
[arXiv:1808.10567 [hep-ex]].

\bibitem{Aushev:2010bq}
T.~Aushev, W.~Bartel, A.~Bondar, J.~Brodzicka, T.~E.~Browder, P.~Chang, Y.~Chao, K.~F.~Chen, J.~Dalseno and A.~Drutskoy, \textit{et al.}
[arXiv:1002.5012 [hep-ex]].

\bibitem{MEGII:2018kmf}
A.~M.~Baldini \textit{et al.} [MEG II],
Eur. Phys. J. C \textbf{78}, no.5, 380 (2018)
[arXiv:1801.04688 [physics.ins-det]].

\bibitem{CMS:2017con}
A.~M.~Sirunyan \textit{et al.} [CMS],
JHEP \textbf{06} (2018), 001.

\bibitem{ATLAS:2019pmk}
G.~Aad \textit{et al.} [ATLAS],
Phys. Lett. B \textbf{800}, 135069 (2020)
[arXiv:1907.06131 [hep-ex]].

\bibitem{ATLAS:2019xlq}
[ATLAS],
ATLAS-CONF-2019-037.

\bibitem{Qin:2017aju}
Q.~Qin, Q.~Li, C.~D.~L\"u, F.~S.~Yu and S.~H.~Zhou,
Eur. Phys. J. C \textbf{78} (2018) no.10, 835
[arXiv:1711.07243 [hep-ph]].

\bibitem{Zee:1980ai}
A.~Zee,
Phys. Lett. B \textbf{93}, 389 (1980)
[erratum: Phys. Lett. B \textbf{95}, 461 (1980)]
doi:10.1016/0370-2693(80)90349-4

\bibitem{Barman:2021xeq}
R.~K.~Barman, R.~Dcruz and A.~Thapa,
JHEP \textbf{03}, 183 (2022)
[arXiv:2112.04523 [hep-ph]].

\bibitem{Herrero-Garcia:2017xdu}
J.~Herrero-Garc\'\i{}a, T.~Ohlsson, S.~Riad and J.~Wir\'en,
JHEP \textbf{04} (2017), 130.

\bibitem{Sabatta:2019nfg}
D.~Sabatta, A.~S.~Cornell, A.~Goyal, M.~Kumar, B.~Mellado and X.~Ruan,
Chin. Phys. C \textbf{44}, no.6, 063103 (2020)
[arXiv:1909.03969 [hep-ph]].

\bibitem{Vicente:2019ykr}
A.~Vicente,
Front. in Phys. \textbf{7}, 174 (2019)
[arXiv:1908.07759 [hep-ph]].

\bibitem{Arganda:2004bz}
E.~Arganda, A.~M.~Curiel, M.~J.~Herrero and D.~Temes,
Phys. Rev. D \textbf{71} (2005), 035011.

\bibitem{Marcano:2019rmk}
X.~Marcano and R.~A.~Morales,
Front. in Phys. \textbf{7}, 228 (2020)
doi:10.3389/fphy.2019.00228
[arXiv:1909.05888 [hep-ph]].

\bibitem{Ilakovac:1999md}
A.~Ilakovac,
Phys. Rev. D \textbf{62} (2000), 036010.

\bibitem{Arganda:2014dta}
E.~Arganda, M.~J.~Herrero, X.~Marcano and C.~Weiland,
Phys. Rev. D \textbf{91} (2015) no.1, 015001.

\bibitem{Arganda:2016zvc}
E.~Arganda, M.~J.~Herrero, X.~Marcano, R.~Morales and A.~Szynkman,
Phys. Rev. D \textbf{95} (2017) no.9, 095029.

\bibitem{Thao:2017qtn}
N.~H.~Thao, L.~T.~Hue, H.~T.~Hung and N.~T.~Xuan,
Nucl. Phys. B \textbf{921}, 159-180 (2017),
arXiv:1703.00896 [hep-ph].

\bibitem{Hernandez-Tome:2020lmh}
G.~Hern\'andez-Tom\'e, J.~I.~Illana and M.~Masip,
Phys. Rev. D \textbf{102}, no.11, 113006 (2020)
[arXiv:2005.11234 [hep-ph]].

\bibitem{Nguyen:2020ehj}
T.~P.~Nguyen, T.~T.~Thuc, D.~T.~Si, T.~T.~Hong and L.~T.~Hue,
PTEP \textbf{2022}, 023 (2022),
arXiv:2011.12181 [hep-ph].

\bibitem{Brignole:2003iv}
A.~Brignole and A.~Rossi,
Phys. Lett. B \textbf{566} (2003), 217-225.

\bibitem{Brignole:2004ah}
A.~Brignole and A.~Rossi,
Nucl. Phys. B \textbf{701} (2004), 3-53.

\bibitem{Diaz-Cruz:2002ezb}
J.~L.~Diaz-Cruz,
JHEP \textbf{05} (2003), 036.

\bibitem{Giang:2012vs}
P.~T.~Giang, L.~T.~Hue, D.~T.~Huong and H.~N.~Long,
Nucl. Phys. B \textbf{864} (2012), 85-112.

\bibitem{Arana-Catania:2013xma}
M.~Arana-Catania, E.~Arganda and M.~J.~Herrero,
JHEP \textbf{09} (2013), 160
[erratum: JHEP \textbf{10} (2015), 192].

\bibitem{Hue:2015fbb}
L.~T. Hue, H.~N. Long, T.~T. Thuc and T.Phong Nguyen,
Nucl. Phys. B \textbf{907} (2016), 37,
arXiv:1512.03266 [hep-ph].

\bibitem{Arganda:2015uca}
E.~Arganda, M.~J.~Herrero, R.~Morales and A.~Szynkman,
JHEP \textbf{03} (2016), 055.

\bibitem{Arganda:2015naa}
E.~Arganda, M.~J.~Herrero, X.~Marcano and C.~Weiland,
Phys. Rev. D \textbf{93} (2016) no.5, 055010.

\bibitem{Zeleny-Mora:2021tym}
M.~Zeleny-Mora, J.~L.~D\'\i{}az-Cruz and O.~F\'elix-Beltr\'an,
[arXiv:2112.08412 [hep-ph]].

\bibitem{Yang:2016hrh}
B.~Yang, J.~Han and N.~Liu,
Phys. Rev. D \textbf{95} (2017) no.3, 035010.



\bibitem{Guo:2016ixx}
H.~K.~Guo, Y.~Y.~Li, T.~Liu, M.~Ramsey-Musolf and J.~Shu,
Phys. Rev. D \textbf{96} (2017) no.11, 115034.

\bibitem{Aoki:2016wyl}
M.~Aoki, S.~Kanemura, K.~Sakurai and H.~Sugiyama,
Phys. Lett. B \textbf{763} (2016), 352-357.

\bibitem{Cheung:2015yga}
K.~Cheung, W.~Y.~Keung and P.~Y.~Tseng,
Phys. Rev. D \textbf{93} (2016) no.1, 015010.

\bibitem{Huitu:2016pwk}
K.~Huitu, V.~Keus, N.~Koivunen and O.~Lebedev,
JHEP \textbf{05} (2016), 026.

\bibitem{Chen:2016lsr}
C.~H.~Chen and T.~Nomura,
Eur. Phys. J. C \textbf{76} (2016) no.6, 353.

\bibitem{Chang:2016ave}
C.~F.~Chang, C.~H.~V.~Chang, C.~S.~Nugroho and T.~C.~Yuan,
Nucl. Phys. B \textbf{910} (2016), 293-308.

\bibitem{Altmannshofer:2015esa}
W.~Altmannshofer, S.~Gori, A.~L.~Kagan, L.~Silvestrini and J.~Zupan,
Phys. Rev. D \textbf{93} (2016) no.3, 031301.

\bibitem{Omura:2015xcg}
Y.~Omura, E.~Senaha and K.~Tobe,
Phys. Rev. D \textbf{94} (2016) no.5, 055019.

\bibitem{Lami:2016mjf}
A.~Lami and P.~Roig,
Phys. Rev. D \textbf{94} (2016) no.5, 056001.

\bibitem{Das:2015zwa}
D.~Das and A.~Kundu,
Phys. Rev. D \textbf{92} (2015) no.1, 015009.

\bibitem{Crivellin:2015mga}
A.~Crivellin, G.~D'Ambrosio and J.~Heeck,
Phys. Rev. Lett. \textbf{114} (2015), 151801.

\bibitem{Campos:2014zaa}
M.~D.~Campos, A.~E.~C\'arcamo Hern\'andez, H.~P\"as and E.~Schumacher,
Phys. Rev. D \textbf{91} (2015) no.11, 116011.

\bibitem{Omura:2015nja}
Y.~Omura, E.~Senaha and K.~Tobe,
JHEP \textbf{05} (2015), 028.

\bibitem{deLima:2015pqa}
L.~de Lima, C.~S.~Machado, R.~D.~Matheus and L.~A.~F.~do Prado,
JHEP \textbf{11} (2015), 074.

\bibitem{Heeck:2014qea}
J.~Heeck, M.~Holthausen, W.~Rodejohann and Y.~Shimizu,
Nucl. Phys. B \textbf{896} (2015), 281-310.

\bibitem{Dorsner:2015mja}
I.~Dor\v{s}ner, S.~Fajfer, A.~Greljo, J.~F.~Kamenik, N.~Ko\v{s}nik and I.~Ni\v{s}and\v{z}ic,
JHEP \textbf{06} (2015), 108.

\bibitem{He:2015rqa}
X.~G.~He, J.~Tandean and Y.~J.~Zheng,
JHEP \textbf{09} (2015), 093.

\bibitem{Dery:2014kxa}
A.~Dery, A.~Efrati, Y.~Nir, Y.~Soreq and V.~Susi\v{c},
Phys. Rev. D \textbf{90} (2014), 115022.

\bibitem{Celis:2013xja}
A.~Celis, V.~Cirigliano and E.~Passemar,
Phys. Rev. D \textbf{89} (2014), 013008.

\bibitem{Falkowski:2013jya}
A.~Falkowski, D.~M.~Straub and A.~Vicente,
JHEP \textbf{05} (2014), 092.

\bibitem{Harnik:2012pb}
R.~Harnik, J.~Kopp and J.~Zupan,
JHEP \textbf{03} (2013), 026.

\bibitem{BhupalDev:2012zg}
P.~S.~Bhupal Dev, R.~Franceschini and R.~N.~Mohapatra,
Phys. Rev. D \textbf{86} (2012), 093010.

\bibitem{Goudelis:2011un}
A.~Goudelis, O.~Lebedev and J.~h.~Park,
Phys. Lett. B \textbf{707} (2012), 369-374.

\bibitem{Diaz-Cruz:1999sns}
J.~L.~Diaz-Cruz and J.~J.~Toscano,
Phys. Rev. D \textbf{62} (2000), 116005.

\bibitem{Korner:1992zk}
J.~G.~Korner, A.~Pilaftsis and K.~Schilcher,
Phys. Rev. D \textbf{47} (1993), 1080-1086.

\bibitem{Pilaftsis:1991ug}
A.~Pilaftsis,
Z. Phys. C \textbf{55} (1992), 275-282.

\bibitem{Pilaftsis:1992st}
A.~Pilaftsis,
Phys. Lett. B \textbf{285} (1992), 68-74.

\bibitem{Blankenburg:2012ex}
G.~Blankenburg, J.~Ellis and G.~Isidori,
Phys. Lett. B \textbf{712} (2012), 386-390.

\bibitem{CarcamoHernandez:2014wdl}
A.~E.~C\'arcamo Hern\'andez, E.~Cata\~no Mur and R.~Martinez,
Phys. Rev. D \textbf{90}, no.7, 073001 (2014)
[arXiv:1407.5217 [hep-ph]].

\bibitem{Jurciukonis:2021izn}
D.~Jur\v{c}iukonis and L.~Lavoura,
JHEP \textbf{03}, 106 (2022)
[arXiv:2107.14207 [hep-ph]].

\bibitem{Aoyama:2020ynm}
T.~Aoyama, N.~Asmussen, M.~Benayoun, J.~Bijnens, T.~Blum, M.~Bruno, I.~Caprini, C.~M.~Carloni Calame, M.~C\`e and G.~Colangelo, \textit{et al.}
Phys. Rept. \textbf{887}, 1-166 (2020)
[arXiv:2006.04822 [hep-ph]].

\bibitem{Aoyama:2012wk}
T.~Aoyama, M.~Hayakawa, T.~Kinoshita and M.~Nio,
Phys. Rev. Lett. \textbf{109}, 111808 (2012)
[arXiv:1205.5370 [hep-ph]].

\bibitem{Aoyama:2019ryr}
T.~Aoyama, T.~Kinoshita and M.~Nio,
Atoms \textbf{7}, no.1, 28 (2019)

\bibitem{Czarnecki:2002nt}
A.~Czarnecki, W.~J.~Marciano and A.~Vainshtein,
Phys. Rev. D \textbf{67}, 073006 (2003)
[erratum: Phys. Rev. D \textbf{73}, 119901 (2006)]
[arXiv:hep-ph/0212229 [hep-ph]].

\bibitem{Gnendiger:2013pva}
C.~Gnendiger, D.~St\"ockinger and H.~St\"ockinger-Kim,
Phys. Rev. D \textbf{88}, 053005 (2013)
[arXiv:1306.5546 [hep-ph]].

\bibitem{Davier:2017zfy}
M.~Davier, A.~Hoecker, B.~Malaescu and Z.~Zhang,
Eur. Phys. J. C \textbf{77}, no.12, 827 (2017)
[arXiv:1706.09436 [hep-ph]].

\bibitem{Keshavarzi:2018mgv}
A.~Keshavarzi, D.~Nomura and T.~Teubner,
Phys. Rev. D \textbf{97}, no.11, 114025 (2018)
[arXiv:1802.02995 [hep-ph]].

\bibitem{Colangelo:2018mtw}
G.~Colangelo, M.~Hoferichter and P.~Stoffer,
JHEP \textbf{02}, 006 (2019)
[arXiv:1810.00007 [hep-ph]].

\bibitem{Hoferichter:2019mqg}
M.~Hoferichter, B.~L.~Hoid and B.~Kubis,
JHEP \textbf{08}, 137 (2019)
[arXiv:1907.01556 [hep-ph]].

\bibitem{Davier:2019can}
M.~Davier, A.~Hoecker, B.~Malaescu and Z.~Zhang,
Eur. Phys. J. C \textbf{80}, no.3, 241 (2020)
[erratum: Eur. Phys. J. C \textbf{80}, no.5, 410 (2020)]
[arXiv:1908.00921 [hep-ph]].

\bibitem{Keshavarzi:2019abf}
A.~Keshavarzi, D.~Nomura and T.~Teubner,
Phys. Rev. D \textbf{101}, no.1, 014029 (2020)
[arXiv:1911.00367 [hep-ph]].

\bibitem{Kurz:2014wya}
A.~Kurz, T.~Liu, P.~Marquard and M.~Steinhauser,
Phys. Lett. B \textbf{734}, 144-147 (2014)
[arXiv:1403.6400 [hep-ph]].

\bibitem{Melnikov:2003xd}
K.~Melnikov and A.~Vainshtein,
Phys. Rev. D \textbf{70}, 113006 (2004)
[arXiv:hep-ph/0312226 [hep-ph]].

\bibitem{Masjuan:2017tvw}
P.~Masjuan and P.~Sanchez-Puertas,
Phys. Rev. D \textbf{95}, no.5, 054026 (2017)
[arXiv:1701.05829 [hep-ph]].

\bibitem{Colangelo:2017fiz}
G.~Colangelo, M.~Hoferichter, M.~Procura and P.~Stoffer,
JHEP \textbf{04}, 161 (2017)
[arXiv:1702.07347 [hep-ph]].

\bibitem{Hoferichter:2018kwz}
M.~Hoferichter, B.~L.~Hoid, B.~Kubis, S.~Leupold and S.~P.~Schneider,
JHEP \textbf{10}, 141 (2018)
[arXiv:1808.04823 [hep-ph]].

\bibitem{Gerardin:2019vio}
A.~G\'erardin, H.~B.~Meyer and A.~Nyffeler,
Phys. Rev. D \textbf{100}, no.3, 034520 (2019)
[arXiv:1903.09471 [hep-lat]].

\bibitem{Bijnens:2019ghy}
J.~Bijnens, N.~Hermansson-Truedsson and A.~Rodr\'\i{}guez-S\'anchez,
Phys. Lett. B \textbf{798}, 134994 (2019)
[arXiv:1908.03331 [hep-ph]].

\bibitem{Colangelo:2019uex}
G.~Colangelo, F.~Hagelstein, M.~Hoferichter, L.~Laub and P.~Stoffer,
JHEP \textbf{03}, 101 (2020)
[arXiv:1910.13432 [hep-ph]].

\bibitem{Blum:2019ugy}
T.~Blum, N.~Christ, M.~Hayakawa, T.~Izubuchi, L.~Jin, C.~Jung and C.~Lehner,
Phys. Rev. Lett. \textbf{124}, no.13, 132002 (2020)
[arXiv:1911.08123 [hep-lat]].

\bibitem{Colangelo:2014qya}
G.~Colangelo, M.~Hoferichter, A.~Nyffeler, M.~Passera and P.~Stoffer,
Phys. Lett. B \textbf{735}, 90-91 (2014)
[arXiv:1403.7512 [hep-ph]].


\bibitem{Pauk:2014rta}
V.~Pauk and M.~Vanderhaeghen,
Eur. Phys. J. C \textbf{74}, no.8, 3008 (2014)
[arXiv:1401.0832 [hep-ph]].


\bibitem{Danilkin:2016hnh}
I.~Danilkin and M.~Vanderhaeghen,
Phys. Rev. D \textbf{95}, no.1, 014019 (2017)
[arXiv:1611.04646 [hep-ph]].

\bibitem{Jegerlehner:2017gek}
F.~Jegerlehner,
Springer Tracts Mod. Phys. \textbf{274}, pp.1-693 (2017)

\bibitem{Knecht:2018sci}
M.~Knecht, S.~Narison, A.~Rabemananjara and D.~Rabetiarivony,
Phys. Lett. B \textbf{787}, 111-123 (2018)
[arXiv:1808.03848 [hep-ph]].

\bibitem{Eichmann:2019bqf}
G.~Eichmann, C.~S.~Fischer and R.~Williams,
Phys. Rev. D \textbf{101}, no.5, 054015 (2020)
[arXiv:1910.06795 [hep-ph]].

\bibitem{Roig:2019reh}
P.~Roig and P.~Sanchez-Puertas,
Phys. Rev. D \textbf{101}, no.7, 074019 (2020)
[arXiv:1910.02881 [hep-ph]].



\bibitem{Muong-2:2021ojo}
B.~Abi \textit{et al.} [Muon g-2],
Phys. Rev. Lett. \textbf{126}, no.14, 141801 (2021)
[arXiv:2104.03281 [hep-ex]].

\bibitem{Muong-2:2006rrc}
G.~W.~Bennett \textit{et al.} [Muon g-2],
Phys. Rev. D \textbf{73} (2006), 072003
[arXiv:hep-ex/0602035 [hep-ex]].


\bibitem{Borsanyi:2020mff}
S.~Borsanyi, Z.~Fodor, J.~N.~Guenther, C.~Hoelbling, S.~D.~Katz, L.~Lellouch, T.~Lippert, K.~Miura, L.~Parato and K.~K.~Szabo, \textit{et al.}
Nature \textbf{593} (2021) no.7857, 51-55
[arXiv:2002.12347 [hep-lat]].

\bibitem{Crivellin:2020zul}
A.~Crivellin, M.~Hoferichter, C.~A.~Manzari and M.~Montull,
Phys. Rev. Lett. \textbf{125}, no.9, 091801 (2020)
[arXiv:2003.04886 [hep-ph]].

\bibitem{Keshavarzi:2020bfy}
A.~Keshavarzi, W.~J.~Marciano, M.~Passera and A.~Sirlin,
Phys. Rev. D \textbf{102}, no.3, 033002 (2020)
[arXiv:2006.12666 [hep-ph]].

\bibitem{Colangelo:2020lcg}
G.~Colangelo, M.~Hoferichter and P.~Stoffer,
Phys. Lett. B \textbf{814}, 136073 (2021)
[arXiv:2010.07943 [hep-ph]].

\bibitem{Morel:2020dww}
L.~Morel, Z.~Yao, P.~Clad\'e and S.~Guellati-Kh\'elifa,
Nature \textbf{588}, no.7836, 61-65 (2020)

\bibitem{Zhang:2021nzv}
Z.~N.~Zhang, H.~B.~Zhang, J.~L.~Yang, S.~M.~Zhao and T.~F.~Feng,
Phys. Rev. D \textbf{103} (2021) no.11, 115015
[arXiv:2105.09799 [hep-ph]].

\bibitem{Baek:2015mea}
S.~Baek and K.~Nishiwaki,
Phys. Rev. D \textbf{93} (2016) no.1, 015002
[arXiv:1509.07410 [hep-ph]].

\bibitem{Foot:1994ym}
R.~Foot, H.~N.~Long and T.~A.~Tran,
Phys. Rev. D \textbf{50} (1994) no.1, R34-R38
[arXiv:hep-ph/9402243 [hep-ph]].

\bibitem{Long:1996rfd}
H.~N.~Long,
Phys. Rev. D \textbf{54} (1996), 4691-4693
[arXiv:hep-ph/9607439 [hep-ph]].

\bibitem{Long:1995ctv}
H.~N.~Long,
Phys. Rev. D \textbf{53} (1996), 437-445
[arXiv:hep-ph/9504274 [hep-ph]].

\bibitem{Hue:2021xap}
L.~T.~Hue, H.~T.~Hung, N.~T.~Tham, H.~N.~Long and T.~P.~Nguyen,
Phys. Rev. D \textbf{104} (2021) no.3, 033007
[arXiv:2104.01840 [hep-ph]].

\bibitem{Zhang:2015csm}
H.~B.~Zhang, T.~F.~Feng, S.~M.~Zhao, Y.~L.~Yan and F.~Sun,
Chin. Phys. C \textbf{41} (2017) no.4, 043106
[arXiv:1511.08979 [hep-ph]].

\bibitem{Nguyen:2018rlb}
T.~P.~Nguyen, T.~T.~Le, T.~T.~Hong and L.~T.~Hue,
Phys. Rev. D \textbf{97} (2018) no.7, 073003
[arXiv:1802.00429 [hep-ph]].

\bibitem{CarcamoHernandez:2020pnh}
A.~E.~C\'arcamo Hern\'andez, L.~T.~Hue, S.~Kovalenko and H.~N.~Long,
Eur. Phys. J. Plus \textbf{136} (2021) no.11, 1158
[arXiv:2001.01748 [hep-ph]].

\bibitem{Hung:2021fzb}
H.~T.~Hung, N.~T.~Tham, T.~T.~Hieu and N.~T.~T.~Hang,
PTEP \textbf{2021} (2021) no.8, 083B01
[arXiv:2103.16018 [hep-ph]].

\bibitem{Montero:1992jk}
J.~C.~Montero, F.~Pisano and V.~Pleitez,
Phys. Rev. D \textbf{47} (1993), 2918-2929
[arXiv:hep-ph/9212271 [hep-ph]].

\bibitem{Singer:1980sw}
M.~Singer, J.~W.~F.~Valle and J.~Schechter,
Phys. Rev. D \textbf{22} (1980), 738

\bibitem{Frampton:1992wt}
P.~H.~Frampton,
Phys. Rev. Lett. \textbf{69} (1992), 2889-2891

\bibitem{Pisano:1992bxx}
F.~Pisano and V.~Pleitez,
Phys. Rev. D \textbf{46} (1992), 410-417
[arXiv:hep-ph/9206242 [hep-ph]].

\bibitem{Lindner:2016bgg}
M.~Lindner, M.~Platscher and F.~S.~Queiroz,
Phys. Rept. \textbf{731} (2018), 1-82
[arXiv:1610.06587 [hep-ph]].

\bibitem{DeJesus:2020yqx}
A.~S.~De Jesus, S.~Kovalenko, F.~S.~Queiroz, C.~Siqueira and K.~Sinha,
Phys. Rev. D \textbf{102} (2020) no.3, 035004
[arXiv:2004.01200 [hep-ph]].

\bibitem{deJesus:2020ngn}
\'A.~S.~de Jesus, S.~Kovalenko, F.~S.~Queiroz, C.~A.~de S.~Pires and Y.~S.~Villamizar,
Phys. Lett. B \textbf{809} (2020), 135689
[arXiv:2003.06440 [hep-ph]].

\bibitem{Hernandez:2021xet}
A.~E.~C.~Hern\'andez, D.~T.~Huong and I.~Schmidt,
Eur. Phys. J. C \textbf{82} (2022) no.1, 63
[arXiv:2109.12118 [hep-ph]].

\bibitem{Hue:2021zyw}
L.~T.~Hue, K.~H.~Phan, T.~P.~Nguyen, H.~N.~Long and H.~T.~Hung,
Eur. Phys. J. C \textbf{82}, no.8, 722 (2022)
[arXiv:2109.06089 [hep-ph]].

\bibitem{Boucenna:2015zwa}
S.~M.~Boucenna, J.~W.~F.~Valle and A.~Vicente,
Phys. Rev. D \textbf{92} (2015) no.5, 053001
[arXiv:1502.07546 [hep-ph]].

\bibitem{Chang:2006aa}
D.~Chang and H.~N.~Long,
Phys. Rev. D \textbf{73} (2006), 053006
[arXiv:hep-ph/0603098 [hep-ph]].

\bibitem{Buras:2012dp}
A.~J.~Buras, F.~De Fazio, J.~Girrbach and M.~V.~Carlucci,
JHEP \textbf{02} (2013), 023
[arXiv:1211.1237 [hep-ph]].

\bibitem{Dreiner:2008tw}
H.~K.~Dreiner, H.~E.~Haber and S.~P.~Martin,
Phys. Rept. \textbf{494} (2010), 1-196
[arXiv:0812.1594 [hep-ph]].

\bibitem{Ninh:2005su}
L.~Ninh and H.~N.~Long,
Phys. Rev. D \textbf{72} (2005), 075004
[arXiv:hep-ph/0507069 [hep-ph]].

\bibitem{Hue:2017lak}
L.~T.~Hue, L.~D.~Ninh, T.~T.~Thuc and N.~T.~T.~Dat,
Eur. Phys. J. C \textbf{78} (2018) no.2, 128
[arXiv:1708.09723 [hep-ph]].

\bibitem{Okada:2016whh}
H.~Okada, N.~Okada, Y.~Orikasa and K.~Yagyu,
Phys. Rev. D \textbf{94} (2016) no.1, 015002
[arXiv:1604.01948 [hep-ph]].

\bibitem{Hung:2019jue}
H.~T.~Hung, T.~T.~Hong, H.~H.~Phuong, H.~L.~T.~Mai and L.~T.~Hue,
Phys. Rev. D \textbf{100} (2019) no.7, 075014
[arXiv:1907.06735 [hep-ph]].

\bibitem{Crivellin:2018qmi}
A.~Crivellin, M.~Hoferichter and P.~Schmidt-Wellenburg,
Phys. Rev. D \textbf{98} (2018) no.11, 113002
[arXiv:1807.11484 [hep-ph]].

\bibitem{Jegerlehner:2009ry}
F.~Jegerlehner and A.~Nyffeler,
Phys. Rept. \textbf{477} (2009), 1-110
[arXiv:0902.3360 [hep-ph]].

\bibitem{Denner:2011mq}
A.~Denner, S.~Heinemeyer, I.~Puljak, D.~Rebuzzi and M.~Spira,
Eur. Phys. J. C \textbf{71} (2011), 1753
[arXiv:1107.5909 [hep-ph]].

\bibitem{ParticleDataGroup:2020ssz}
P.~A.~Zyla \textit{et al.} [Particle Data Group],
PTEP \textbf{2020} (2020) no.8, 083C01

\bibitem{T2K:2019bcf}
K.~Abe \textit{et al.} [T2K],
Nature \textbf{580} (2020) no.7803, 339-344
[erratum: Nature \textbf{583} (2020) no.7814, E16]
[arXiv:1910.03887 [hep-ex]].

\bibitem{ParticleDataGroup:2018ovx}
M.~Tanabashi \textit{et al.} [Particle Data Group],
Phys. Rev. D \textbf{98} (2018) no.3, 030001

\bibitem{tHooft:1972tcz}
G.~'t Hooft and M.~J.~G.~Veltman,
Nucl. Phys. B \textbf{44} (1972), 189-213

\bibitem{Denner:2005nn}
A.~Denner and S.~Dittmaier,
Nucl. Phys. B \textbf{734} (2006), 62-115
[arXiv:hep-ph/0509141 [hep-ph]].

\bibitem{Hahn:1998yk}
T.~Hahn and M.~Perez-Victoria,
Comput. Phys. Commun. \textbf{118} (1999), 153-165
[arXiv:hep-ph/9807565 [hep-ph]].

\bibitem{Phan:2016ouz}
K.~H.~Phan, H.~T.~Hung and L.~T.~Hue,
PTEP \textbf{2016} (2016) no.11, 113B03
[arXiv:1605.07164 [hep-ph]].



\bibitem{T2K:2019bcf}
K.~Abe \textit{et al.} [T2K],
Nature \textbf{580}, no.7803, 339-344 (2020)
[erratum: Nature \textbf{583}, no.7814, E16 (2020)]
[arXiv:1910.03887 [hep-ex]].

\bibitem{Enomoto:2019mzl}
K.~Enomoto, S.~Kanemura, K.~Sakurai and H.~Sugiyama,
Phys. Rev. D \textbf{100}, no.1, 015044 (2019)
[arXiv:1904.07039 [hep-ph]].

\bibitem{Camara:2020efq}
H.~B.~Camara, R.~G.~Felipe and F.~R.~Joaquim,
JHEP \textbf{05}, 021 (2021)
[arXiv:2012.04557 [hep-ph]].

\bibitem{Nebot:2007bc}
M.~Nebot, J.~F.~Oliver, D.~Palao and A.~Santamaria,
Phys. Rev. D \textbf{77}, 093013 (2008)
[arXiv:0711.0483 [hep-ph]].

\bibitem{Fernandez-Martinez:2016lgt}
E.~Fernandez-Martinez, J.~Hernandez-Garcia and J.~Lopez-Pavon,
JHEP \textbf{08}, 033 (2016)
[arXiv:1605.08774 [hep-ph]].

\bibitem{Agostinho:2017wfs}
N.~R.~Agostinho, G.~C.~Branco, P.~M.~F.~Pereira, M.~N.~Rebelo and J.~I.~Silva-Marcos,
Eur. Phys. J. C \textbf{78}, no.11, 895 (2018)
[arXiv:1711.06229 [hep-ph]].

\bibitem{Coutinho:2019aiy}
A.~M.~Coutinho, A.~Crivellin and C.~A.~Manzari,
Phys. Rev. Lett. \textbf{125}, no.7, 071802 (2020)
[arXiv:1912.08823 [hep-ph]].

\bibitem{Manzari:2020eum}
C.~A.~Manzari, A.~M.~Coutinho and A.~Crivellin,
PoS \textbf{LHCP2020}, 242 (2021)
[arXiv:2009.03877 [hep-ph]].

	
\end{thebibliography}
\end{document}